\documentclass[preprint,twocolumn]{aastex61}
\usepackage{amsmath}
\usepackage{mathtools}
\usepackage{natbib}
\usepackage{threeparttable}
\usepackage{booktabs}
\usepackage{bibentry}

\def\eg{{\it e.g.,}}
\def\ie{{\rm i.e.,}}

\def\micron{{$\mu$m}}

\newcommand{\be}{\begin{equation}}
\newcommand{\ee}{\end{equation}}

\newcommand{\crash}{{\sc CRASH}}
\newcommand{\omegafac}{{Omega}}

\begin{document}
\nobibliography*
\title{Laboratory Photoionization Fronts in Nitrogen Gas: A Numerical Feasibility and Parameter Study}
\author{William J. Gray}
\affil{CLASP, College of Engineering, University of Michigan, 2455 Hayward St., Ann Arbor, Michigan 48109, USA}
\author{P. A. Keiter}
\affil{CLASP, College of Engineering, University of Michigan, 2455 Hayward St., Ann Arbor, Michigan 48109, USA}
\author{H. Lefevre}
\affil{CLASP, College of Engineering, University of Michigan, 2455 Hayward St., Ann Arbor, Michigan 48109, USA}
\author{C. R. Patterson}
\affil{CLASP, College of Engineering, University of Michigan, 2455 Hayward St., Ann Arbor, Michigan 48109, USA}
\author{J. S. Davis}
\affil{CLASP, College of Engineering, University of Michigan, 2455 Hayward St., Ann Arbor, Michigan 48109, USA}
\author{B. van Der Holst}
\affil{CLASP, College of Engineering, University of Michigan, 2455 Hayward St., Ann Arbor, Michigan 48109, USA}
\author{K. G. Powell}
\affil{Department of Aerospace Engineering, University of Michigan, Ann Arbor, Michigan, 48109, USA}
\author{R. P. Drake}
\affil{CLASP, College of Engineering, University of Michigan, 2455 Hayward St., Ann Arbor, Michigan 48109, USA}


\begin{abstract}
	Photoionization fronts play a dominant role in many astrophysical situations, but remain difficult to achieve in a laboratory experiment. We present the results from a computational parameter study evaluating the feasibility of the photoionization experiment presented in the design paper by \bibentry{Drake2016} in which a photoionization front is generated in a nitrogen medium . The nitrogen gas density and the Planckian radiation temperature of the x-ray source define each simulation. Simulations modeled experiments in which the x-ray flux is generated by a laser-heated gold foil, suitable for experiments using many kJ of laser energy, and experiments in which the flux is generated by a ``z-pinch'' device, which implodes a cylindrical shell of conducting wires. The models are run using \crash, our block-adaptive-mesh code for multi-material radiation hydrodynamics. The radiative transfer model uses multi-group, flux-limited diffusion with thirty radiation groups. In addition, electron heat conduction is modeled using a single-group, flux-limited diffusion. In the theory, a photoionization front can exist only when the ratios of the electron recombination rate to the photoionization rate and the electron impact ionization rate to the recombination rate lie in certain ranges. These ratios are computed for several ionization states of nitrogen. Photoionization fronts are found to exist for laser driven models with moderate nitrogen densities ($\sim$10$^{21}$ cm$^{-3}$) and radiation temperatures above 90 eV. For ``z-pinch'' driven models, lower nitrogen densities are preferred ($<$10$^{21}$ cm$^{-3}$). We conclude that the proposed experiments are likely to generate photoionization fronts.
\end{abstract}

\section{Introduction}
Photoionization fronts are important drivers of change at all scales in the universe \citep{Robertson2010}. The cosmic Dark Ages, so named because no sources of emission existed, began once all the hydrogen recombined some 370,000 years after the Big Bang. During the Dark Ages the first dense structures begin to form \citep{Press1974} made mostly of dark matter with virial masses of M$_{vir}\sim$10$^{6}$ M$_{\odot}$ \citep{Greif2015}. It is in these dark matter ``minihalos' that the first stars formed with masses of M$_{*}\sim$100 M$_{\odot}$\citep{Schaerer2002}. These Population III stars are much more compact, have higher surface temperatures, and produce many more UV photons than present day stars \citep{Tumlinson2000}.

In addition to forming in ``minihalos'', the first galaxies hosted a substantial population of the first stars. The first galaxies formed as these ``minihalos'' merged together. Once a dark matter halo reaches a mass of $\sim$10$^{8}$ M$_{\odot}$ atomic cooling becomes efficient and a population of stars forms \citep{Wise2007,Wise2008,Greif2008,Greif2010,Bromm2011}. In addition to heating and ionizing the surrounding gas, these first stars created photoionization fronts that were fundamental in creating structure within these galaxies. Finally, in combination with black holes, these first stars began to reionize the universe, ending the cosmic Dark Ages.

Photoionization fronts also play an important role in the present day universe. Massive stars M$_{*}>20 $M$_{\odot}$, with their high surface temperatures, produce a large number of ionizing photons. These photons ionize the region around the stars, creating nebulae known as H II regions \citep[\eg][]{Franco2000,Williams2000,Mackey2016,Gvaramadze2017}. These H II regions have important effects on the surrounding gas, from destroying giant molecular clouds to affecting future star formation \citep{McKee2007}. Photoionization fronts are also important in the transition from ``pre-planetary'' to ``planetary'' nebula during the late stage evolution of mid-mass stars ($\sim$1-8 M$_{\odot}$) \citep{Tafoya2013,Gledhill2015,Planck2015}.

To date there have been only a handful of laboratory experiments that might have generated a photoionization front. The experiments by Willi and collaborators \citep{Afsharrad1994,Hoarty1999a,Hoarty1999b,Willi2000} irradiated low-density triacrylate foams with a soft x-ray source to study the motion of the heat front where ionization occurred, without much focus on the ionization mechanism. The x-ray source was generated by laser-irradiation of a ``burnthrough'' gold foil. Using x-ray spectroscopy and radiography these authors inferred density and temperature profiles through the foam target with results that suggested a supersonic ionization front. Similarly, \cite{Zhang2010} irradiated the interior of a gold cylinder to generate an x-ray source and create an ionization front in a low-density plastic (C$_{8}$H$_{8}$) foam. Like the Willi experiments, these authors used time resolved x-ray radiography to measure ionization and shock positions in their foams.  However, \cite{Drake2016} showed that these experiments did not actually generate photoionization fronts, but instead produced heat fronts in which the energy was carried by electron heat conduction.

\cite{Drake2016} outlined the theoretical requirements for a true laboratory photoionization front experiment. They considered experiments in which an x-ray source of finite diameter irradiates nitrogen gas held at high pressure. Their study suggested, from analytical calculations, that a radiation source temperature of T$_{\rm R}$=100 eV and an initial nitrogen pressure of 10 atmospheres is sufficient to generate a  photoionization front. As mentioned in \cite{Drake2016} the primary photoabsorption mechanism is photoelectric, that is, the freed electron has an energy equal to the photon energy minus the ionization energy. In astrophysical systems, stars providing the ionizing photons have temperatures of a few eV and only the photons in the high energy tail drive the photoionizations. Therefore, the freed electrons have an energy of only a few eV. In contrast, laboratory sources have temperatures much higher than the characteristic ionization energies, which translates to much higher electron energies. However, as stressed in \cite{Drake2016}, even if these laboratory experiments fail to precisely match the astrophysical conditions, a structured experiment of a photoionization front is a meaningful first step.

In the present paper, we improve upon the work presented in \cite{Drake2016} in two ways. We develop and show results from an improved analytic atomic model, using three rather than two ionization states. But we note that neither of the analytic models accounts for energy transport by electron heat conduction, for two-dimensional effects, or for the hydrodynamic response of the ionized gas. To determine whether these mechanisms affect the viability of photoionization front experiments, or change the range over which they are possible, one must do computational simulations. Here we present results from a suite of two dimensional radiation hydrodynamic simulations in order to study the propagation of the radiation front in potential experiments. We conduct a parameter study that varies the source radiation temperatures and nitrogen pressure in order to determine the optimal conditions to form a photoionization front. We consider two experimental platforms, one where the x-ray source is generated from a laser heated foil and one where the source is generated from the implosion of current carrying wires.

The structure of the paper is as follows. In \S2 we present the theoretical background of ionization fronts as heat fronts and results from the expanded three level theoretical atomic model. In \S3 we give the framework for the simulations and our initial conditions. \S4 presents important numerical considerations and caveats. In \S5 we give the results of our simulations and in \S6 we give the summary and conclusion.

\section{Theory}
\subsection{Ionization Fronts}
\label{sec:IonFronts}

\begin{figure*}
\begin{center}
\includegraphics[trim=0.0mm 0.0mm 0.0mm 0.0mm, clip, scale=0.65]{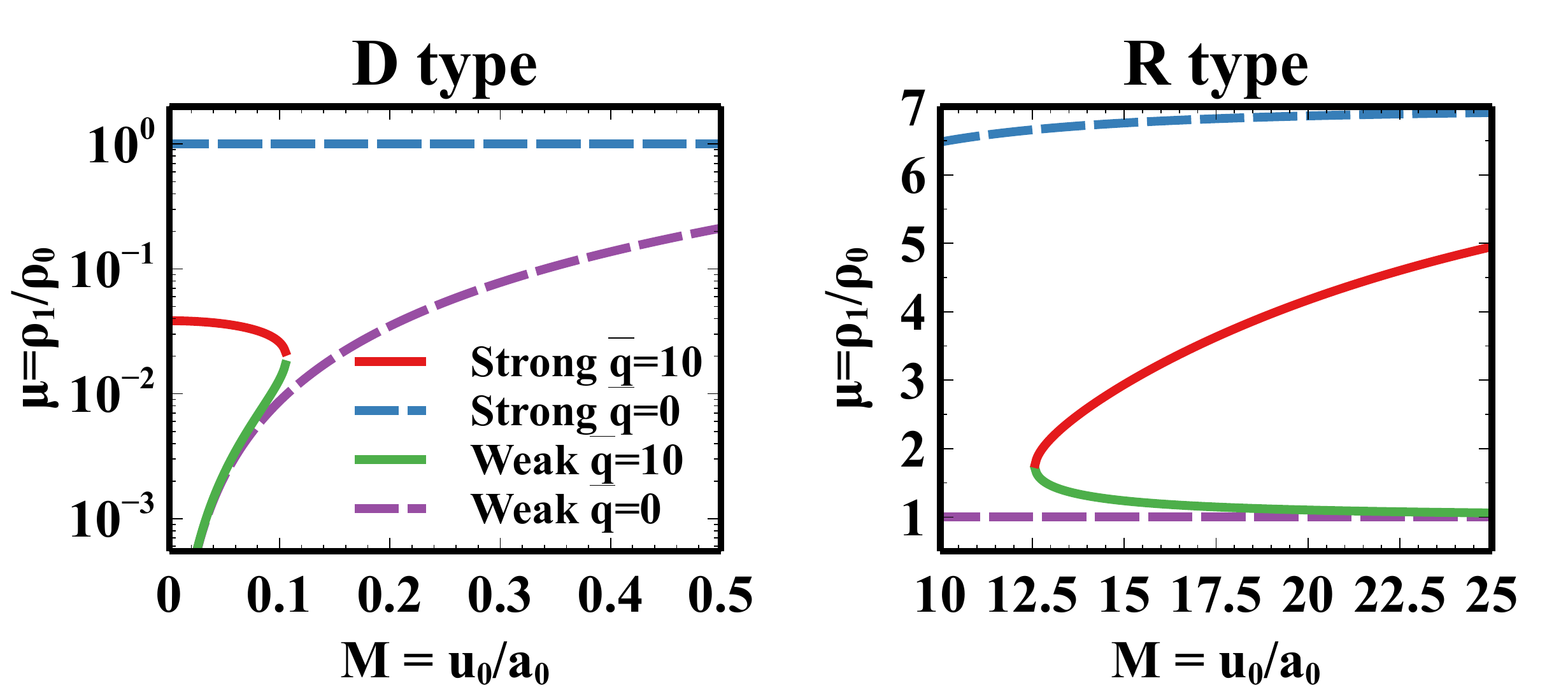} 
\caption{Density ratios for {\it Left Panel:} Subsonic (note the logarithmic scale) and {\it Right Panel:} supersonic flows. The $y$-axis gives the density ratio $\mu$ and $x$ axis gives the Mach number. The solid lines shows the $\overline{q}$=0 solutions while the dashed lines shows the $\overline{q}$=10 solutions. }
\label{fig:mach}
\end{center}
\end{figure*}

As discussed in \cite{Drake2016}, photoionization events deposit considerable amounts of energy at the front. In addition, the photoionized medium downstream from the front has a much higher temperature than the material upstream. This allows the photoionization front to be described as a heat front. Astrophysical applications of heat fronts are considered in \citep[][]{Kahn1954,Goldsworthy1958,Goldsworthy1961,Axford1961,Osterbrock2006}. A similar analysis is useful in the laboratory system considered here.

We start by moving to a frame of reference that follows the front. A planar, steady front is assumed and we label the conditions upstream of the front with the subscript $0$ and those downstream by $1$. The conservation of mass, momentum, and energy are then written as
\begin{equation}
\rho_0u_0 = \rho_1u_1 \label{eqn:mass},
\end{equation}
\begin{equation}
P_0 + \rho_0u_0^2 = P_1 + \rho_1u_1^2 \label{eqn:mom},
\end{equation}
\begin{equation}
\begin{split}
\left(\frac{1}{2}\rho_0u_0^2 +\frac{\gamma}{\gamma-1}P_0 - q^2\rho_0 \right)u_0 =\\ \left(\frac{1}{2}\rho_1u_1^2 +\frac{\gamma}{\gamma-1}P_1 \right)u_1, \label{eqn:energy}
\end{split}
\end{equation}
where $\rho$ is the mass density, $u$ is the velocity, P is the pressure, $\gamma$ is the adiabatic index of the gas, and $q$ measures the internal energy deposited in the gas during photoionization and has units of velocity for convenience. Following \cite{Osterbrock2006}, $q$, having units of velocity, is defined as
\begin{equation}
\phi_i\left(\frac{1}{2}m_iq^2\right)=\int_{\nu_0}^{\infty}\frac{\pi F_{\nu}}{h\nu}(h\nu-h\nu_0)d\nu,
\end{equation}
where $m_i$ is the mean mass of the ionized gas, $\phi_i$ is the ionizing photon flux arriving at the front, and $F_{\nu}$ is the spectral energy flux of the source, a function of frequency $\nu$. It is with this definition that we are allowed to write Equation~\ref{eqn:energy}. If the front were instead generated by a diffusive heat front, then the incoming energy flux would depend on the gradient of the temperature. This would lead to very different structure at the front compared to that generated by a photoionization front \citep{Drake2018}.

Using Equation~\ref{eqn:mass} and Equation~\ref{eqn:mom} to simplify Equation~\ref{eqn:energy}, one finds
\begin{align}
\rho_0\left(u_0^2+a_0^2\right) &= \rho_1\left(u_1^2+a_1^2\right) \label{eqn:smom} \\
\frac{1}{2}\left(u_1^2-u_0^2\right)&+\frac{\gamma}{\gamma-1}\left(a_1^2-a_0^2\right) = q^2 \label{eqn:sen}
\end{align}

In order to simplify this further and arrive at an equation for the density ratio, we solve Equation~\ref{eqn:smom} for a$_1$,
\begin{equation}
a^2_1 = \frac{1}{\mu^2}\left( u_0^2(\mu-1)+\mu a_0^2 \right),
\end{equation}
where $\mu$=$\rho_1$/$\rho_0$. Substituting this into Equation~\ref{eqn:sen} and making use of Equation~\ref{eqn:mass} we arrive at
\begin{equation}
\label{eqn:musq}
\mu^2(u_0^2+ka_0^2+2q^2)-k(u_0^2+a_0^2)\mu + u_0^2(k-1) = 0,
\end{equation}
where $k = 2\gamma/(\gamma-1)$. Following \cite{Drake2016}, the isothermal sound speed, $a^2=p/\rho$, upstream of the front is independent of the front and is useful in parameterizing the flow. Dividing through by a$_0^2$, Equation~\ref{eqn:musq} becomes
\begin{equation}
\mu^2(M^2+k+2\bar{q}^2)-k(M^2+1)\mu + M^2(k-1) = 0,
\end{equation}
where $M$ is the upstream Mach number and $\bar{q}$=$q$/a$_0$. This can be solved for two real, positive roots,
\begin{equation}
\begin{aligned}
\label{eqn:mu}
\mu = \frac{\rho_1}{\rho_0} = \frac{Ak\pm\sqrt{k^2A^2-4BM^2(k-1)}}{2B},
\end{aligned}
\end{equation}
where $A=M^2+1$ and $B=M^2+k+2\bar{q}^2$.

Equation~\ref{eqn:mu} is now dependent only on known quantities, the upstream Mach number, $M$, the ratio of specific heats $\gamma$, and $\overline{q}$. As mentioned above, $q$ is related to the excess energy deposited in the front by photons. $q$ and, conversely, $\bar{q}$ can be estimated for astrophysical environments by assuming an idealized hydrogen nebula surrounding an O-type star. The upstream sound speed is roughly a$_0\sim$10 km/s for T=10,000 K. The energy deposited by the photons implies approximately $q^{2}\sim 2k_{B}T_s/\overline{M}$, where $T_s$ is the temperature of the source, $k_{B}$ is Boltzmann's constant, and $\overline{M}$ is the average atomic mass. For a 52,000 K O-type star, $q\sim$ 30 km/s and $\overline{q}$=3. For laboratory experiments, \cite{Drake2016} estimated $\overline{q}\sim$10.

Figure~\ref{fig:mach} shows the density ratios for two cases for Equation~\ref{eqn:mu}. The $\bar{q}$=0 case represents the classic jump conditions without radiation while the $\bar{q}$=10 case represents the laboratory conditions discussed above. A $\gamma$=4/3 is assumed for both cases.

Figure~\ref{fig:mach} also shows the four classic classifications of shock fronts. The D-type, or ``dense'' front, are fronts that move subsonically with respect to the upstream gas and are shown in the left panel. Strong D-type fronts are supersonic rarefactions where the density drops by some amount at the front. Weak D-type fronts are fronts with subsonic rarefactions. In the absence of radiation, the density is essentially unchanged across the front. With radiation, there is some modest heating which leads to lower densities behind the front. The right panel shows the R-type, or ``rarified'' fronts. The strong R-type is a modified shock wave where the radiation lowers the compression in the shock. The weak R-type is the classic, simple ionization front where the photons are absorbed in a thin layer near the front but leave the density largely unchanged. In nature, strong R-type fronts are unstable and are not seen.

It should be noted that the heat-front type evolves as the front moves outward. As discussed by \cite{Osterbrock2006}, if we consider a uniform neutral medium and instantaneously turn on a large source of photons, the heat front generated will start as a weak R-type, that is, a photoionization front. Eventually the flux of ionizing photons drops low enough, due to geometric reduction and recombinations that an R-type front is unsustainable and a shock is formed followed by a strong D-type front.

\subsection{Three stage atom}
\label{sec:threelevel}
\begin{figure}
\begin{center}
\includegraphics[trim=0.0mm 0.0mm 0.0mm 0.0mm, clip, width=0.95\columnwidth]{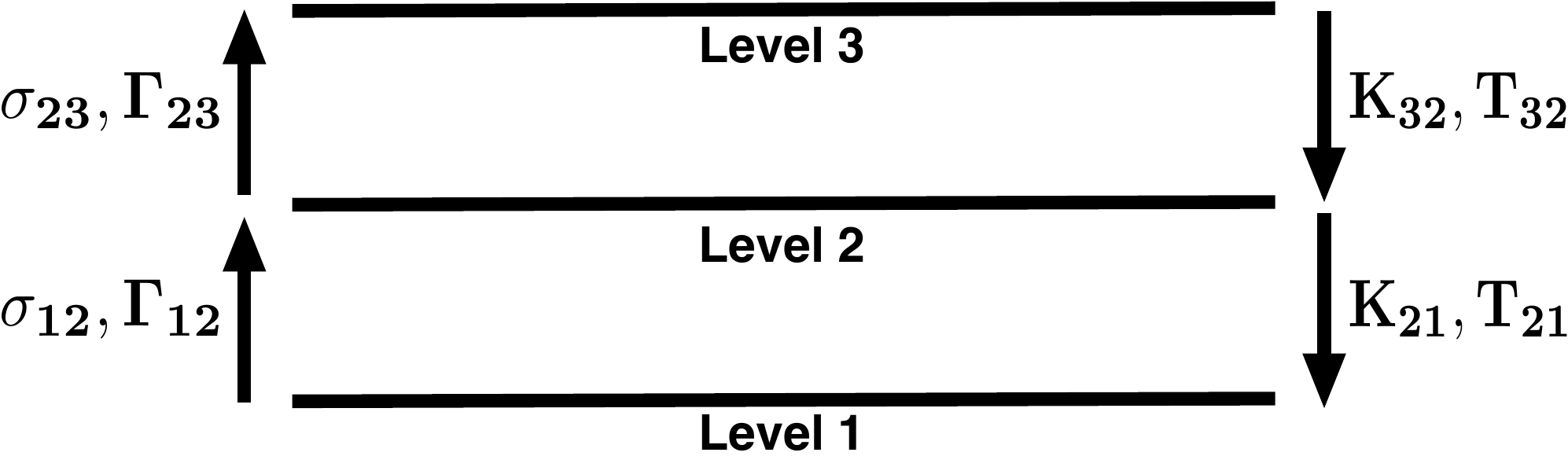}
\caption{Schematic view of the three level atom. $\sigma_{12}$ and $\sigma_{23}$ represent electron impact ionization rate coefficients, K$_{12}$ and K$_{32}$ represent electron recombination rate coefficients, T$_{23}$ and T$_{21}$ represent three body electron recombination rate coefficients, and $\Gamma_{12}$ and $\Gamma_{23}$ represent photoionization rates. }
\label{fig:model}
\end{center}
\end{figure}

\begin{figure*}
\begin{center}
\includegraphics[trim=0.0mm 0.0mm 0.0mm 0.0mm, clip, scale=0.65]{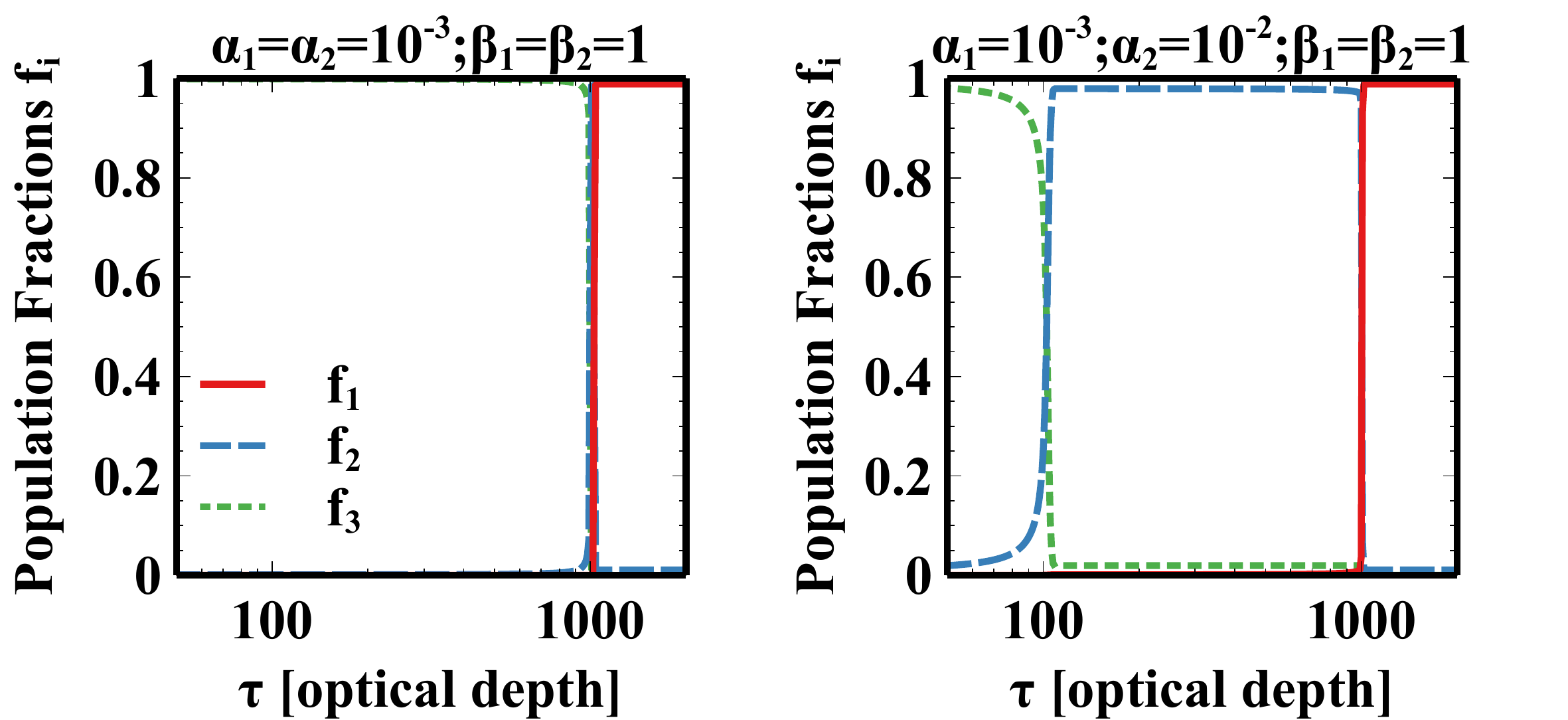}
\caption{Three level model atom results. The left panel shows results for $\alpha_{1}$=$\alpha_{2}$=10$^{-3}$ whereas the right panel shows $\alpha_{1}$=10$^{-3}$ and $\alpha_{2}$=10$^{-2}$.  $\beta_1=\beta_2$=1 and Z$_i$=1 for both panels. Here the $x$-axis shows the initial optical depth for the lowest state considered.}
\label{fig:modelfrac}
\end{center}
\end{figure*}

The extent of a photoionized region depends on the flux of photoionizing photons and the recombination rate. \cite{Drake2016} used a two-level model atom to study photoionization fronts and found the structure of the ionization front is strongly determined by the ratio of the recombination rate to the photoionization rate and the ratio of the electron impact ionization rate to the  recombination rate. Here we expand this model to three levels in order to evaluate the effects of multi-level atoms and additional energy levels on photoionization front structures.

Consider a model atom with three distinct energy levels with increasing ionization energies. Between three adjacent ion stages, three atomic processes are considered: ionization by electron collision, electron recombination, and photoionization. Figure~\ref{fig:model} shows a schematic view of this model.  We consider a planar model and that photoionization is the only mechanism that decreases the photon flux, \ie{} geometric effects are ignored. The level population for each ionization state is determined by, along with equations for charge conservation and mass conservation
\begin{align}
\label{eqn:dn1}
\frac{dn_1}{dt} = &-\sigma_{12} n_1 n_e - \Gamma_{\ell,12} n_1 \\
                  &+ K_{21} n_2 n_e + T_{21} n_2 n_e n_e \nonumber \\
\label{eqn:dn2}
\frac{dn_2}{dt} = &-\left(\frac{dn_1}{dt}+\frac{dn_3}{dt}\right) \\
\label{eqn:dn3}
\frac{dn_3}{dt} = &\sigma_{23} n_2 n_e + \Gamma_{\ell,23} n_2 \\
                  &-K_{32} n_3 n_e - T_{32} n_3 n_e n_e \nonumber \\
\label{eqn:nE}
n_e = &Z_{1}n_{1}+Z_{2}n_{2}+Z_{3}n_{3} \\
\label{eqn:nT}
n_{T} = &n_{1}+n_{2}+n_{3},
\end{align}
where $\sigma_{12}$ and $\sigma_{23}$ are the electron collision ionization rate coefficients for levels 1$\rightarrow$2 and levels 2$\rightarrow$3 respectively, $\Gamma_{\ell,12}$ and $\Gamma_{\ell,23}$ are the local photoionization rates, K$_{21}$ and K$_{32}$ are the recombination rate coefficients, and T$_{21}$ and T$_{32}$ are the three body recombination rate coefficients. n$_e$ is the electron number density while n$_1$, n$_2$, and n$_3$ are the number densities in each level. Z$_{1}$, Z$_{2}$, and Z$_{3}$ are the degree of ionization for each level.

Equations~\ref{eqn:dn1} and \ref{eqn:dn3} show the recombination terms in the traditional form, respecting that the leading behavior involves the product of two densities for two-body processes and three densities for three-body processes. In this case, the two-body coefficients are independent of density as one might hope, but the three-body coefficients retain some density dependence. Informed by the determination in \cite{Drake2016} that three-body recombination is not dominant in laboratory systems that produce a photoionization front, we choose to slightly overstate the total rate of recombination as follows. We rewrite the recombination terms between states $i$ and $j$ as R$_{ji}$=$(K_{ji} + T_{ji} Z_3 n_T) n_j n_e$, in which $T_{ji}$ is evaluated at an electron density of $Z_3 n_T$. This gives a conservative result, in the sense that conditions found to produce an ionization front of a certain size will in actuality produce one at least that large.

Equations~\ref{eqn:dn1},\ref{eqn:dn3}-\ref{eqn:nT} can be rewritten as, assuming steady state,
\begin{align}
\label{eqn:dn1r}
-&f_{1}(\alpha_{\ell,1}(\beta_{1}-1)f_{e} + 1) + \alpha_{\ell,1}f_{2}f_{e}=0 \\
&f_{2}(\alpha_{\ell,2}(\beta_{2}-1)f_{e} + 1) - \alpha_{\ell,2}f_{3}f_{e}=0 \\
&f_{1}+f_{2}+f_{3}-1=0
\end{align}
where,
\begin{align}
\label{eqn:beta1}
\beta_{1}&=\sigma_{12}/R_{21}+1 \\
\label{eqn:beta2}
\beta_{2}&=\sigma_{23}/R_{32}+1 \\
\label{eqn:alpha1}
\alpha_{\ell,1}&=R_{21}n_{T}/\Gamma_{\ell,12} \\
\label{eqn:alpha2}
\alpha_{\ell,2}&=R_{32}n_{T}/\Gamma_{\ell,23} \\
\end{align}
$f_{e}=Z_{1}f_{1}+Z_{2}f_{2}+Z_{3}f_{3}$ and f$_i$=n$_i$/n$_T$.

In order to complete our system of equations we need to account for the decrease in photon flux due to photoionization events. The spatial evolution of the radiation flux is defined as
\begin{equation}
\label{eqn:dFR}
\frac{dF_{R}}{dx} = -\sigma_{1} F_R n_1 - \sigma_{2} F_R n_2,
\end{equation}
where $\sigma_{1}$ and $\sigma_{2}$ are the spectrally averaged photoionization cross sections (\eg{} see Eqn~\ref{eqn:sigmabar} below) for ionization state n$_1$ and n$_2$ respectively. This can be recast in terms of the initial optical depth $\tau=n_T\sqrt{\sigma_1^2+\sigma_2^2}$. The local radiation flux is related to the source radiation flux by $F_{R\ell}$=$F_{R}$g($\tau$) which recasts Eqn.~\ref{eqn:dFR} as
\begin{equation}
\frac{dg(\tau)}{d\tau}=-\frac{(f_{1}\sigma_1+f_{2}\sigma_1)g(\tau)}{\sqrt{\sigma_1^2+\sigma_2^2}}.
\end{equation}
Finally, in order to relate the local photoionization rate to the rate at the source, we define $\alpha_{\ell,i}$ as $\alpha_{\ell,i} = \alpha_i/g(\tau)$.

These equations are solved for values relevant in laboratory experiments, $\alpha\sim$10$^{-2}$-10$^{-4}$, $\beta_1=\beta_2$=1, $\sigma_1=\sigma_2$, and Z$_i$=1 and are shown in Figure~\ref{fig:modelfrac}. An iterative approach is taken in solving the above equations. First, using an initial estimate for g($\tau$), the ionization state fractions are updated using standard root solving algorithms. These are then used to update g($\tau$) using a common ODE solver.  The left panel of Figure~\ref{fig:modelfrac} shows results for $\alpha_i$=10$^{-4}$ while the right shows $\alpha_1$=10$^{-4}$ and $\alpha_2$=10$^{-3}$. For equal values of $\alpha$ the system quickly transitions from a fully ionized state to a fully neutral state, that is, there is essentially no region where the singly ionized state is present. This is in contrast to the unequal $\alpha$ case. Here there is an appreciable region where the singly ionized state is dominant. This suggests that for photoionization experiments with multi-level atoms, one could see many ionization fronts, one for each ionization state. In practice, this depends on the ionization energy of each state or the electron configuration of the atom.

\section{Model Framework and Initial Conditions}

\begin{figure}
\begin{center}
\includegraphics[trim=0.0mm 0.0mm 0.0mm 0.0mm, clip, width=0.49\columnwidth,angle=270]{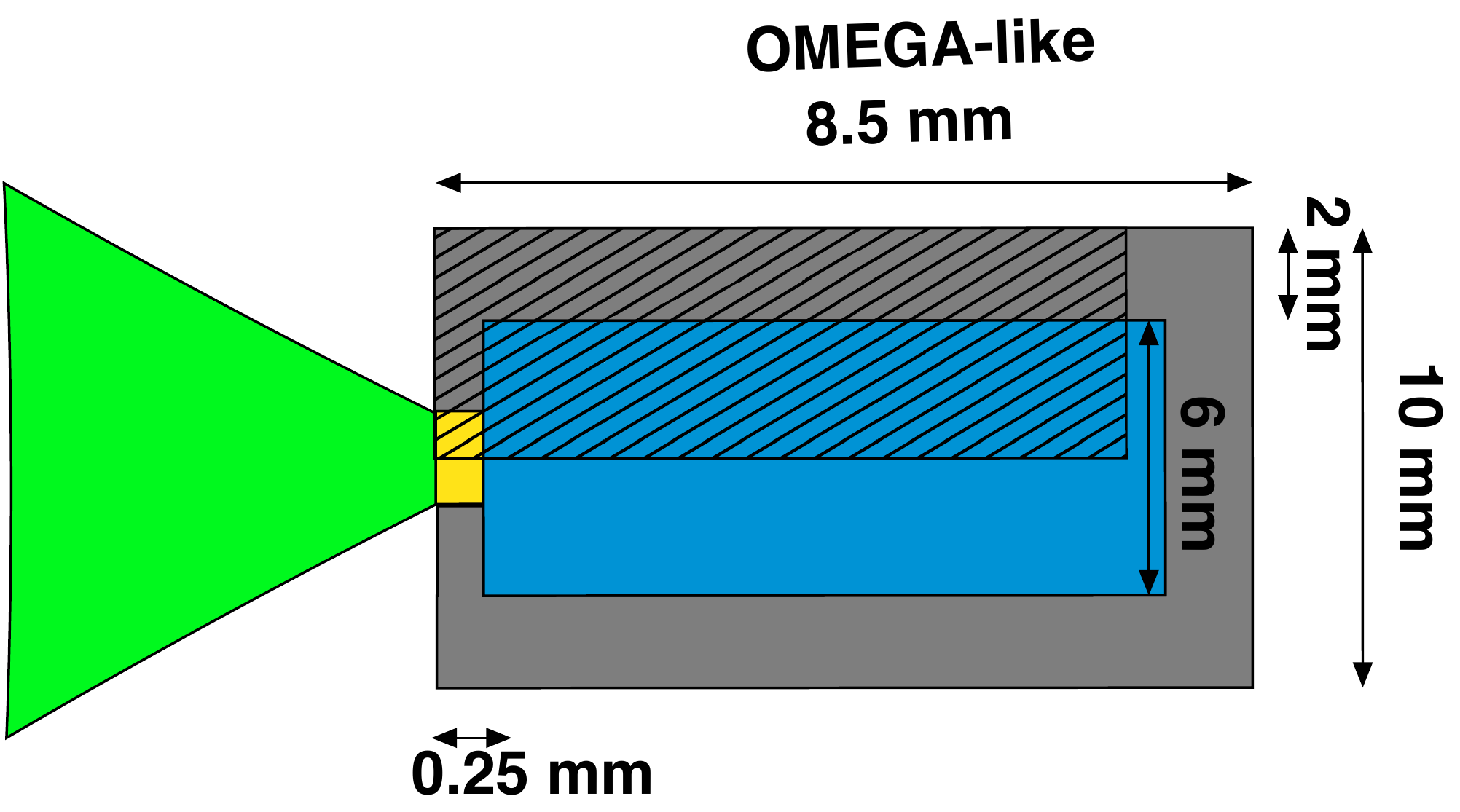}
\caption{Schematic representation of the laser driven experiment design. The gray region represents the acrylic tube. The blue region represents the gas cell containing the nitrogen gas. The gold region represents the domain where the gold foil and support structure is found. The green region denotes the lasers impinging on the gold foil. Experiment dimensions are also given. The crosshatched region shows the simulated domain.}
\label{fig:setup}
\end{center}
\end{figure}

\begin{figure}
\begin{center}
\includegraphics[trim=0.0mm 0.0mm 0.0mm 0.0mm, clip, width=0.49\columnwidth,angle=270]{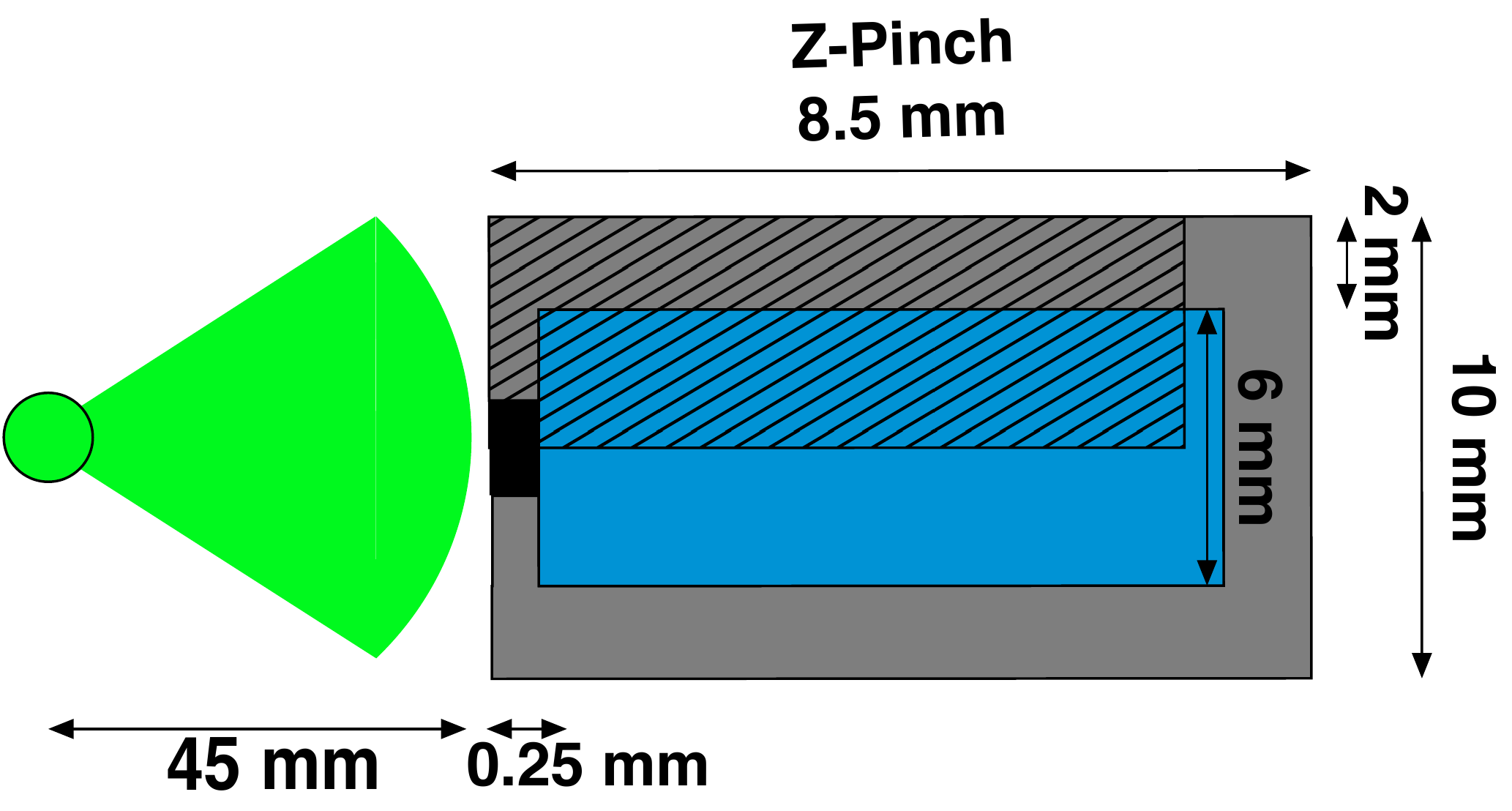}
\caption{Schematic representation of the ``z-pinch'' experiment design. The gray region represents the acrylic tube. The blue region represents the gas cell containing the nitrogen gas. The black region represents the domain where the support structure is found. The green region denotes the irradiating x-rays. The crosshatched region shows the simulated domain.}
\label{fig:setupZ}
\end{center}
\end{figure}

\begin{figure*}
\begin{center}
\includegraphics[trim=0.0mm 0.0mm 0.0mm 0.0mm, clip, scale=0.7 ,angle=0]{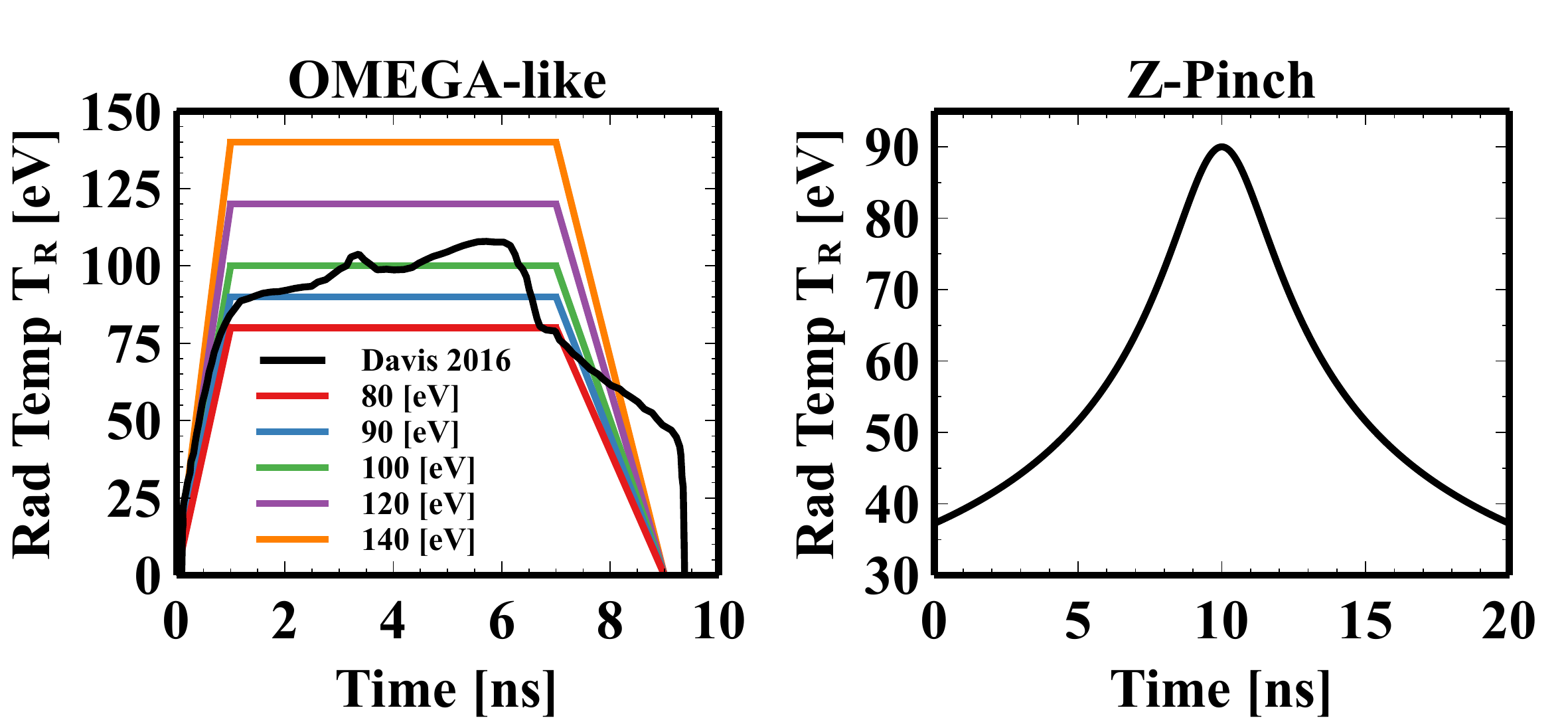}
\caption{ {\it Left Panel: } Radiation temperature profile in eV for the laser heating shots. The black line shows the 0.5 $\mu$m x-ray emission temperature profile reported by \cite{Davis2016}. All other lines show the time dependent radiation temperature profiles used in the simulations; the legend gives the peak temperature used here. {\it Right Panel: } Radiation temperature profile for the ``z-pinch'' shots. This is derived from the x-ray power presented in \cite{Rochau2014}. }
\label{fig:laserprofile}
\end{center}
\end{figure*}

All the simulations presented here were performed using \crash, our adaptive mesh code for multi-material radiation hydrodynamics \citep{vanderHolst2011}. The mesh is continuously refined and derefined using the Block-Adaptive Tree Library (BATL)\citep{Toth2012}. Different materials are distinguished using level sets that are designed to prevent material mixing. These levels sets are advected along with the flow. Material properties, such as average ionization state and opacity, are stored on invertible tables that are read in at runtime. For nitrogen, we employ {\sc Propaceos}-generated tables (see Appendix A of \cite{MacFarlane2006}). In generating these tables, {\sc Propaceos} utilizes the {\sc ATBASE} suite of atomic physics codes in order to generate the appropriate atomic transition rates. For all other material we use \crash\ generated tables. Radiation is treated using multi-group flux-limited diffusion with a total of thirty groups logarithmically spanning an energy range of 0.1-2.0$\times$10$^{4}$ eV. Electron heat conduction is modeled using a single-group, flux-limited diffusion with a constant flux limiter equal to 0.06.

Every simulation employs a base resolution of 43.75 $\mu$m with up to five levels of AMR, for a fine resolution of 1.4 $\mu$m. Refinement is based on changes in density and pressure, which is sufficient to track the developing front. In addition, the equation of state for each material is provided as invertible tables read in at runtime. Again, we employ {\sc Propaceos} generated tables for nitrogen and \crash\ generated tables for all other materials.

Although there are many methods of generating x-ray flux to drive experiments, only two are considered here. These are laser heating of a gold foil and flux generated by a ``z-pinch'' machine. In the laser-heating scenario, heating a gold foil irradiated by a set of lasers generates the x-ray flux. In ``z-pinch'' machines, imploding a current carrying material generates the x-ray flux. Here we use the \omegafac\, facility \citep{Boehly1995} as a model for laser heating scenarios and the Z Astrophysical Plasma Properties (ZAPP) platform \citep{Rochau2014} on the Sandia National Laboratory Z facility as a model for ``z-pinch'' machines.

\subsection{\omegafac\, Setup}

Each simulation is run in two dimensions with Cartesian coordinates with the x coordinate ranging between 0 and 7 mm and the $y$ coordinate ranging between 0 and 3.5 mm. Figure~\ref{fig:setup} schematically shows the target design. The outer tube, shown in gray, is made of an acrylic plastic. The gold region contains the gold foil and supporting structures, which once heated provides the radiation drive into the nitrogen gas. The impinging lasers are represented by the green triangle. The nitrogen gas is held within the target by a gas window consisting of a thin polyimide film supported by a stainless steel grid. The crosshatch pattern shows the simulated region. There is small region below the outer tube and to the left of the gas window. This is a vacuum region in the experiment and is modeled here as a void region filled with xenon gas at very low density.

For simplicity and to save on computation time, we make two simplifications in modeling this experiment. First, we ignore the gold foil. The resolution requirements are reduced since the gold foil is very thin compared to the size of the nitrogen gas cell. The gas window, however, is included and is modeled as a 5 $\mu$m piece of acrylic plastic. Secondly, the laser heating and subsequent emission from the gold foil is modeled as a radiation temperature boundary condition.

\cite{Davis2016} studied the emission from laser-heated gold foils. These results serve as the basis for our radiation temperature boundary condition. These authors measured the x-ray emission using the DANTE x-ray spectrometer and derived the radiation temperature as a function of time by fitting the DANTE spectra to a Planckian profile. The left panel of Figure~\ref{fig:laserprofile} shows the results of this experiment for a 0.5 $\mu$m gold foil. Also shown is the parameterized radiation temperature boundary conditions used in the \omegafac-like simulations. The radiation temperature rises linearly with time, reaching the peak by 1 ns. The temperature is then held fixed for the next 5 ns. The temperature then falls linearly in time to zero over the last 2 ns. Finally, since the gold foil is located in front of the entrance to the gas cell, the radiation boundary condition is only implemented over the entrance to the gas cell. For all other hydrodynamic variables a reflecting boundary condition is used. Reflecting boundary conditions are used for the upper and lower $y$ boundaries, while outflow conditions are used for the right $x$ boundary.

We report results from a total of twenty \omegafac-like simulations. The incident radiation temperature, and consequently the incident laser intensity, is varied between 80 and 140 eV. This can be achieved by varying the intensity of the incident lasers. The pressure in the gas cell is varied between five and forty atmospheres. This corresponds to a number density of nitrogen atoms between 2.5$\times$10$^{20}$ and 2.0$\times$10$^{21}$ cm$^{-3}$. Each simulation is run for a total simulation time of 2 ns, which allows the radiation drive to peak and emit at its peak temperature for 1 ns. In the discussion presented below, the nominal run is a model with a nitrogen pressure of ten atmospheres, or equivalently a density of 5.0$\times$10$^{20}$cm$^{-3}$, and a drive temperature of 100 eV, matching the ideal setup as described in \cite{Drake2016}.

\subsection{ZAPP Setup}

The right panel of Figure~\ref{fig:laserprofile} shows the x-ray drive shape as a function of time derived using the x-ray power profile from the ZAPP platform, as show in Figure 3 of \cite{Rochau2014}. The drive is mirrored around the peak in order to obtain a symmetric profile. This profile is then fit with a Cauchy profile, providing the drive shape as a function of time used for the radiation boundary condition.

Figure 6 of \cite{Rochau2014} gives insight into the spectrum and intensity of the drive. A typical photoionization experiment on the Sandia Z machine is placed between 45 and 100 mm from the pinch, which generates a near Planckian emission source, \eg\ \cite{Foord2004}. The drive spectrum is well fit by a single temperature of T$_{\rm R}\sim93$ eV. For a circular source of radius R$_{s}$, the radiation flux is geometrically reduced with distance, $D$, from the source. The total flux as a function of distance from the sources is given by, where we assume that the distance $D$ is some small multiple of the source size
\begin{equation}
	F(D)= \sigma T_R^4 \left(\frac{R_{s}^2}{R_{s}^2+D^2}\right) \sim \sigma T_R^4 \left(\frac{R_s}{D}\right)^2,
\end{equation}
where $\sigma$ is the Stefan-Boltzmann constant, and  $T_R$ is the radiation temperature.
Thus, the radiation at $D$ is reduced relative to the blackbody flux at the source by a factor of $(R_{s}/D)^2$, defining a ``geometrical reduction factor'' at $D$. For a radiation temperature of T$_{\rm R}=93$ eV, the surface flux is 7.7 TW/cm$^{2}$. The measured flux at a distance of $D$=45 mm from the ZAPP source is 1.7 TW/cm$^{2}$, which gives a  reduction factor of 0.22. The source size, R$_s$, is defined by the left $y$ boundary and is set to 3.5 mm. Therefore, to achieve the proper flux at the opening of the target, the left $x$ boundary is set to a distance of 7.45 mm. In addition, the radiation temperature source extends over the entire boundary instead of over a small portion as in the \omegafac-like simulations described above. Schematically, this is shown in Figure~\ref{fig:setupZ}.

We report a total of six $z$-pinch simulations, which vary the nitrogen pressure (equivalently the nitrogen density). This is in contrast to the \omegafac-like simulation where the peak drive temperature is also varied. This is due to the ZAPP platform target placement and pinch design. Each simulation is run for 10 ns, much longer than the \omegafac-like simulation, which allows the radiation drive to reach its peak value. Finally, a wider range of nitrogen densities are studied with the ZAPP setup versus the \omegafac-like setup.

\subsection{Experiment Diagnostics}
Because the simulations reported here are intended to be directly relevant to experiments, it is sensible to discuss anticipated observables and diagnostics. The primary diagnostic is anticipated to be streaked x-ray absorption spectroscopy. This allows for measurement of the velocity of the front in addition to the plasma parameters, \ie\ the electron temperature, electron number density, and average charge.

X-ray absorption spectroscopy provides information on the density and temperature of the target based upon the photons absorbed by the target. This requires a stable x-ray source having a relatively flat spectrum. Such a source is known as a backlighter. The backlighter provides a 300 ps, 2-4 keV source for the absorption spectroscopy using a 1\% Argon dopant in the nitrogen gas \citep{Keiter2016}.

A secondary measurement will use the streaked optical pyrometer (SOP) diagnostic \citep{Miller2007}. This device is typically used to measure the self-emission in laser-driven shocks. There are various options, under consideration, to couple this line of sight to an interferometer which might provide further evidence regarding electron density and front speed. Additional details of the experiment results and diagnostics are given in \cite{Lefevre2018}.

\section{Caveats and Numerical Considerations}

The code evaluates radiative transfer in the  multi-group, flux-limited, diffusion (FLD) approximation. By design FLD assumes that the radiative flux is, $\bf{F_{R}}$$\sim$D$\nabla E_{R}$, where $F_{R}$ is the radiation energy flux, $E_{R}$ is the radiation energy density, and D is a diffusion coefficient that includes the flux limiter that ensures that the radiation propagation speed does not exceed the speed of light \cite[\eg][]{Minerbo1978,Levermore1981,vanderHolst2011}. FLD, is therefore, unable to correctly model optically thin regimes in complex geometries \citep{Krumholz2007}. FLD is also unable to resolve complex structure, such as shadows, as this model acts to fill in regions with low $E_R$. Many of these concerns are mitigated here by the simplified simulation geometry and limiting the analysis to line outs near the bottom $x$-axis where gradients in the radiation energy are minimized. The radiative transfer in optically thick matter, as for example within the neutral nitrogen gas, is well approximated by a diffusion approximation. However, this is not true for the ionized gas where the ionizing photons are free streaming. The flux limiter employed by \crash\, is the so-called square-root limiter, which has the form \citep{vanderHolst2011}
\begin{equation}
D_{g} = \frac{c}{\sqrt{\left(3\kappa_{Rg}\right)^2+\frac{|\nabla E_g|^2}{E_g^2}}},
\end{equation}
where $D_g$ is the diffusion coefficient for radiation group $g$, $c$ is the speed of light, $\kappa_{Rg}$ is the Rosseland mean opacity for group $g$, and $E_g$ is the radiation energy of group $g$. When the radiation length scale, $L_R=E_g/|\nabla E_g|$, the diffusion coefficient simplifies to the diffusive limit. For small radiation length scales, $D_g=c|E_g|/|\nabla E_g|$, and the radiation propagates at the speed of light which is appropriate in the optically thin regime.

It is also important to note that although FLD has the above limitations, it is energy conserving. This fundamental property ensures that shock fronts and heat fronts are properly captured during the course of the simulation. It also ensures that the amount of radiative energy imposed by the boundary condition is approximately the same as experienced by the target in the experiment.

Another concern is the evolution of the ionization states in the nitrogen gas. \crash\, is a general use multi-material hydrodynamics code that requires certain restrictions on the equation of state. For a given material important thermodynamic quantities, \ie\, electron temperature, pressures, and particle number density, are stored as invertible lookup tables and used as the simulation progresses. These tables also include the ionization state of the gas, pre-computing the mean ionization of the material. Therefore, \crash\, does not solve the rate equations to compute the ionization state of the gas as the simulation runs, but simply uses values corresponding to local thermodynamic equilibrium. It is certain the mean ionization derived from a collisional radiative model would differ from that obtained from the lookup table. We therefore take the following results as an approximation to the actual ionization state and treat them as an upper limit \cite[\eg][]{Klapisch2011}.

Finally, we note the presence of the gas window will alter the exact x-ray spectrum seen by the nitrogen. Current experimental designs employ a gas window in which the stainless steel grid covers 15\% of the opening. The presence of the gas window by itself does not alter the spectrum, but the plasma ablating from the grid can selectively absorb the photons and alter the spectrum. This plasma will occupy a small fraction of the window area. Similarly, once the polyimide film has burned through it too will have only a very small effect on the source spectrum.

\section{Results}

\subsection{Hydrodynamics}

\begin{figure*}
\begin{center}
\includegraphics[trim=0.0mm 0.0mm 0.0mm 0.0mm, clip, width=0.85\textwidth,angle=0]{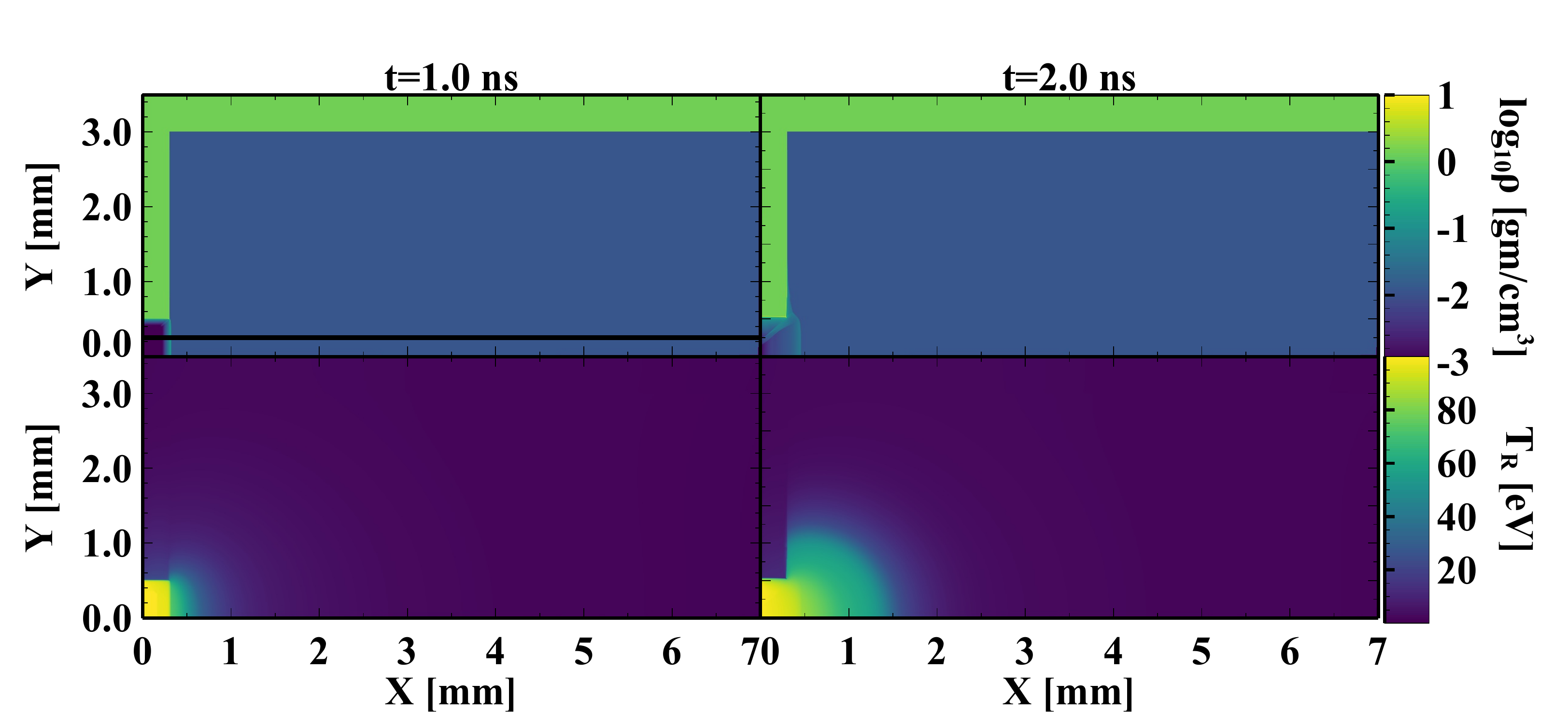}
\caption{Two-dimensional results from the nominal \omegafac-like run with the {\it Top Row:} logarithic density in units of gm/cm$^{-3}$ and the {\it Bottom Row:} radiation temperature in units of eV. The left column shows the results at t=1 ns while the right column shows the results at t=2 ns. The black line shows the line-out path used for Figure~\ref{fig:lineout}. }
\label{fig:2DPlotsNorm}
\end{center}
\end{figure*}

\begin{figure*}
\begin{center}
\includegraphics[trim=0.0mm 0.0mm 0.0mm 0.0mm, clip, width=0.85\textwidth,angle=0]{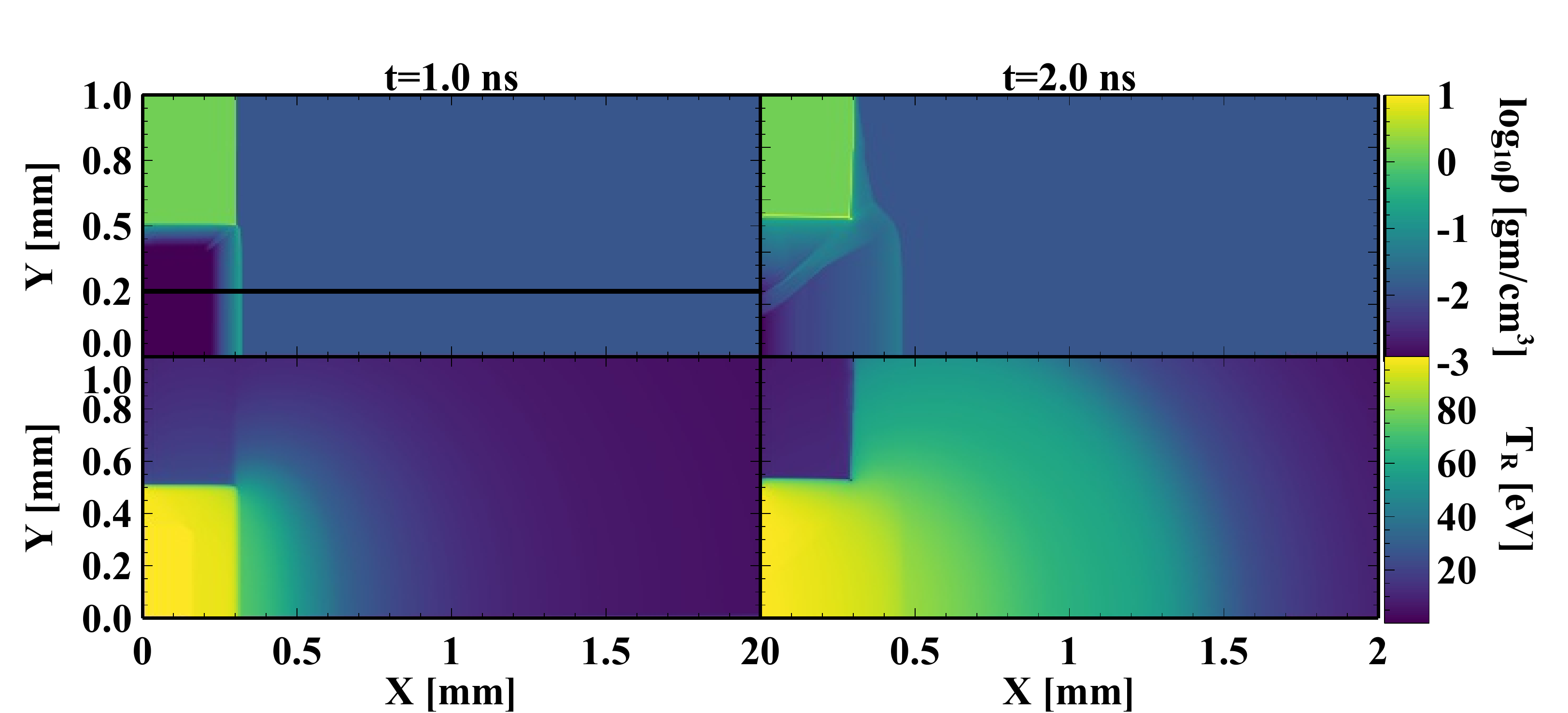}
\caption{Zoom in of Figure~\ref{fig:2DPlotsNorm} near the entrance to the nitrogen gas cell. Note that the density peak seen near x=0.5 mm is due to the plastic gas window. The nitrogen gas is density is essentially unchanged.}
\label{fig:2DPlotsZoom}
\end{center}
\end{figure*}

\begin{figure*}
\begin{center}
\includegraphics[trim=0.0mm 0.0mm 0.0mm 0.0mm, clip, scale=0.70 ,angle=0]{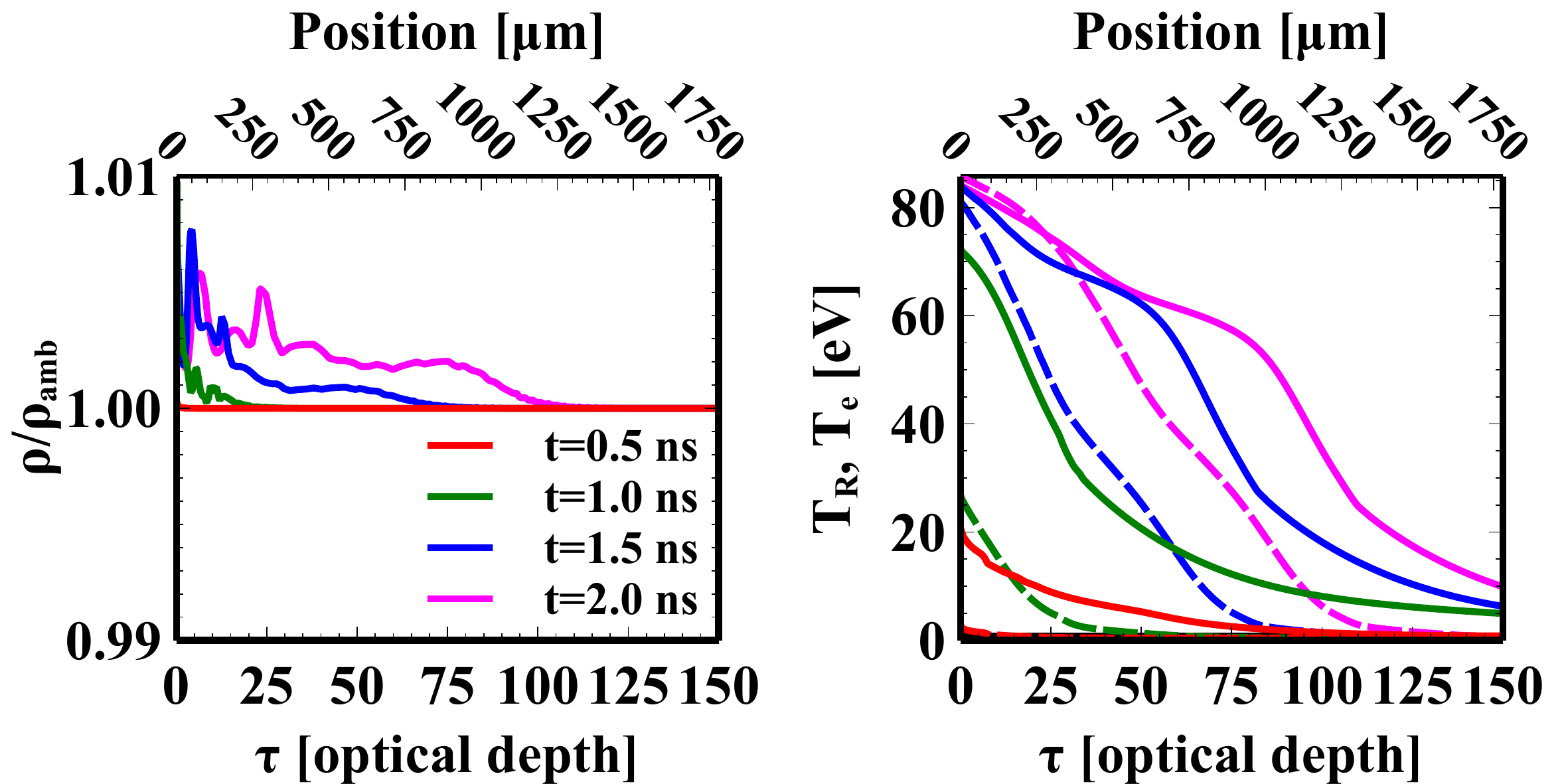}
\caption{ Density ratio (left panel) and temperature (right panel) profiles for the nominal \omegafac-like model. Radiation temperatures are shown as the solid lines while the dashed lines show the electron temperature. The bottom $x$-axis is the initial optical depth through the nitrogen gas assuming no change in ionization state while the top $x$-axis shows the x position along the line-out.}
\label{fig:lineout}
\end{center}
\end{figure*}

Figure~\ref{fig:2DPlotsNorm} shows the two-dimensional density and radiation temperature plots for the nominal model (T$_{\rm R}$=100 eV, 10 atm of nitrogen, \omegafac-like setup). The top row shows the mass density while the bottom row shows the radiation temperature. The first column shows the results at 1 ns and the second column at 2 ns. Very little density evolution is seen between these two times. On the other hand, there is clear evolution of the radiation temperature. To highlight the slight density evolution, Figure~\ref{fig:2DPlotsZoom} zooms in on the region near the entrance to the nitrogen gas cell. At 1 ns the density shows some compression at the interface between the nitrogen gas cell and the void region at 0.3 mm. This is mostly from the destruction of plastic gas window. The nitrogen gas density is essentially unchanged. The radiation temperature, however, has penetrated much further along the gas cell.

In order to illustrate this more clearly, Figure~\ref{fig:lineout} shows line-out profiles through the nitrogen gas cell whose path is shown as the black line in Figure~\ref{fig:2DPlotsNorm}. The left panel of Figure~\ref{fig:lineout} shows the density ratio, defined as the density divided by the ambient density, as a function of time. The $x$-axis is the optical depth through the system assuming that the gas is composed of only neutral nitrogen and does not take into account any evolution of the ionization state of the gas. Once we have accounted for the plastic gas window, there is essentially no evolution in the nitrogen gas density at all times. In fact, even at 2 ns, the maximum compression in the nitrogen gas is only a few percent at an optical depth of $\tau\sim50$. The total, initial optical depth through the nitrogen gas is $\tau\sim$700, which is much larger than optical depths studied here. This is to allow experimental diagnostics to study the front as a function of time as it moves through the gas cell.

The right panel of Figure~\ref{fig:lineout} shows the line-out profile for both the electron temperature (dashed lines) and the radiation temperature (solid lines). The difference in evolution between the two temperatures is striking. At early times, $t=0.5$ ns, the rise in radiation temperature can be seen with almost no change in electron temperature. However, by $t=1.5$ ns, both the radiation and electron temperatures are $\sim$90 eV near the entrance to the nitrogen gas. The difference between these two temperatures may be explained by looking at the photoionization timescales. A simple estimate for the photon flux near the source is
\begin{equation}
\label{eqn:fluxgamma}
F_{\gamma} = \frac{\sigma_{SB}T^{4}_{s}}{2.7k_{B}T_{s}}= 2.36\times10^{23}T_{s}^3,
\end{equation}
where $\sigma_{SB}$ is the Stefan-Boltzmann constant, k$_{B}$ is Boltzmann's constant and T$_{s}$ is the source temperature in units of eV. The photoionization timescale is then
\begin{equation}
\tau_{PI} = \frac{10^{-12}}{\bar{\sigma}F_{\gamma}},
\end{equation}
where $\tau_{PI}$ is the photoionization timescale in units of ps and $\bar{\sigma}$ is the photoionization cross section and is on the order of 10$^{-18}$ cm$^{2}$. At $t=0.5$ ns as the source is ramping up and has a value of T$_{s}$=50 eV and gives a photoionization timescale of 30 ps. At $t=1.5$ ns after the source has reached its peak the timescale has decreased to 6 picoseconds.

The temperature profiles at late times have $T_{R}$ well above $T_e$. The ionizing photons from the source that penetrate the shocked plastic have begun to ionize the gas in front of the shock. This creates a slight knee in the radiation temperature profile at $t=1.5-2$ ns.

The results shown in Figure ~\ref{fig:lineout} can also be used to estimate the photoionization front velocity. The front velocity is estimated by
\begin{equation}
v_{f} = \frac{F_{\gamma}}{n} = \frac{D\nabla{E_{R}}}{{2.7T_{R}n}},
\end{equation}
where, F$_{\gamma}$ is the photon flux, $n$ is the particle number density, and T$_{R}$ is the local radiation temperature in units of eV. At $t=1.5$ ns, the front is located at $\sim$840 \micron\ while at $t=2.0$ ns the front is at $\sim$ 1100 \micron. This gives a front velocity of $\sim$500 \micron/ ns. The radiation temperature near the front is $\sim$40 eV, which by using the flux given by Equation~\ref{eqn:fluxgamma} gives a front velocity of $\sim$300 \micron/ns, consistent with the simulation result. The photoionization front therefore travels $\sim$3\micron\ in a single photoionization timescale. Thus, near the leading edge of the front one would expect the nitrogen to become multiply ionized over a short distance.

It is also important to note the separation between electron and radiation temperatures in Figure~\ref{fig:lineout}. As mentioned in \cite{Drake2016} electron heat conduction can significantly contribute to the overall energy flux when $T_e\sim$$T_{\rm R}$ and would produce nearly identical $T_{\rm R}$ and $T_e$ profiles. Electron heat conduction is only important when the electron temperature is greater than the radiation temperature and when the particle density is greater than 10$^{21}$ cm$^{-3}$. The results shown here for our nominal model show that electron heat conduction does not play an important role here.

\subsection{Photoionization Front}

\begin{figure}
\begin{center}
\includegraphics[trim=0.0mm 0.0mm 0.0mm 0.0mm, clip, width=0.95\columnwidth,angle=0]{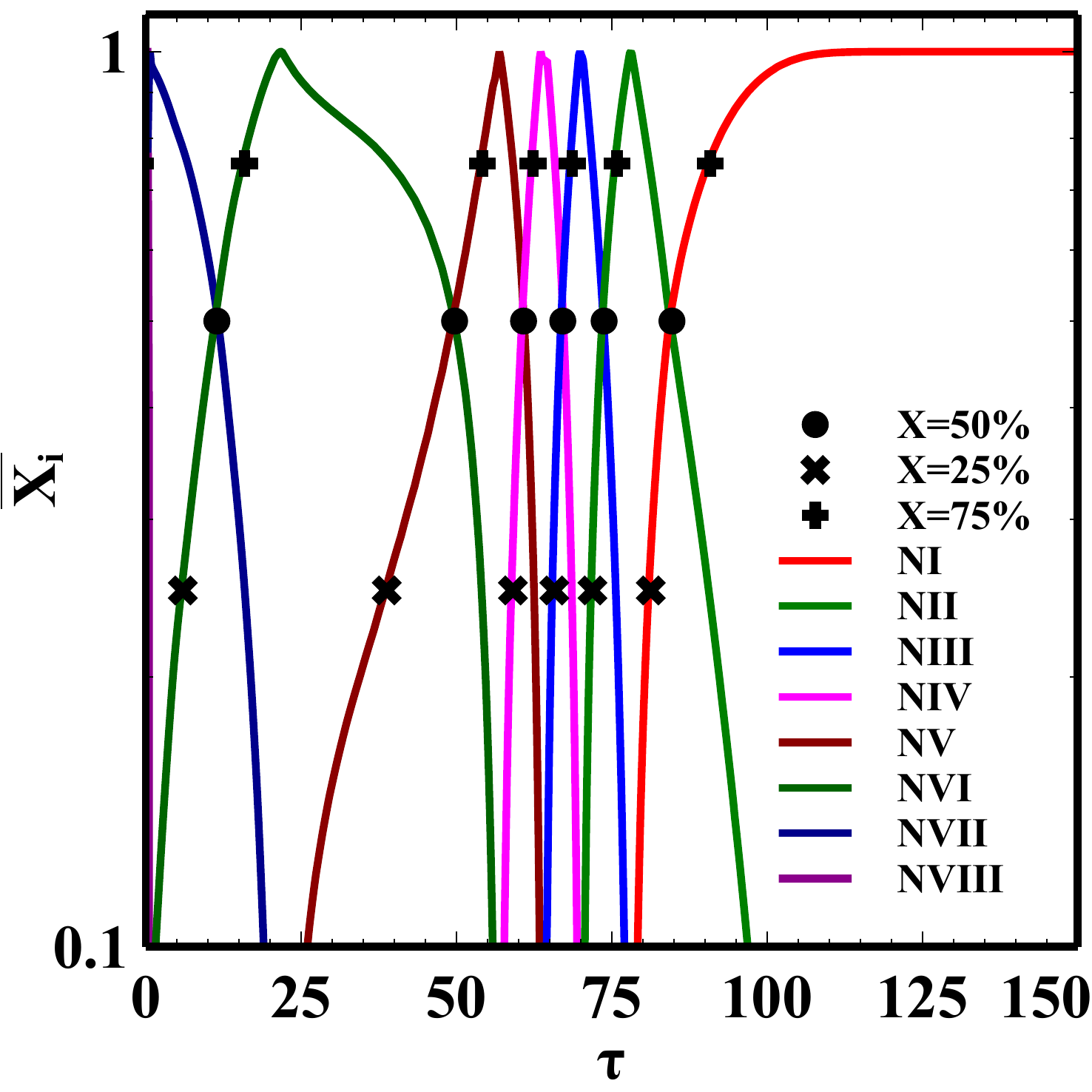}
\caption{Normalized ionization states for our nominal model at t=1.5 ns. The $x$-axis is the initial optical depth through the original, neutral nitrogen gas and the $y$-axis is the normalized ionization fractions. The black symbols give the 25, 50, and 75\% ionization fraction values for each state. The nominal location of the front for each state is the 50\% value.}
\label{fig:SpecTau}
\end{center}
\end{figure}

\begin{figure*}
\begin{center}
\includegraphics[trim=0.0cm 0.0cm 0.0cm 0.0cm,clip,scale=0.6]{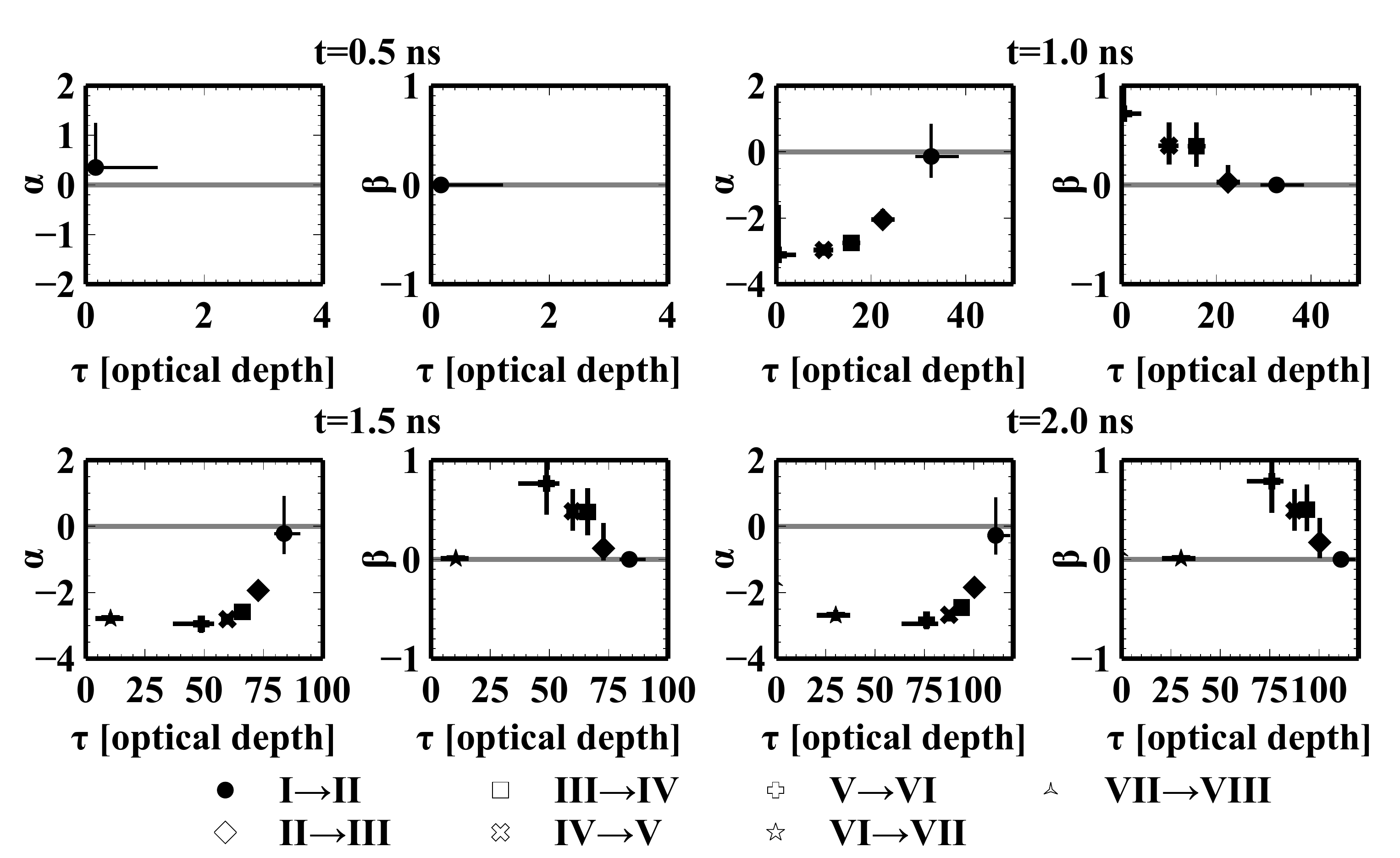}
\caption{Plots of $\alpha$ and $\beta$ for the nominal model at {\it Top Left}: t=0.5 ns, {\it Top Right}: t=1.0 ns, {\it  Bottom Left}: t=1.5 ns, and {\it Bottom Right}: t=2.0 ns. Each symbol represents values between ionization states as shown in the legend. The symbols represent the X$_i$/X$_{i,max}$=50\% values and the error bars correspond to the 25\% and 75\% values. Gray lines at $\alpha$=1 and $\beta$=1 are shown to guide the eye.}
\label{fig:FidTauPlot}
\end{center}
\end{figure*}

\begin{figure*}
\begin{center}
\includegraphics[trim=0.0mm 0.0mm 0.0mm 0.0mm,clip,scale=0.5]{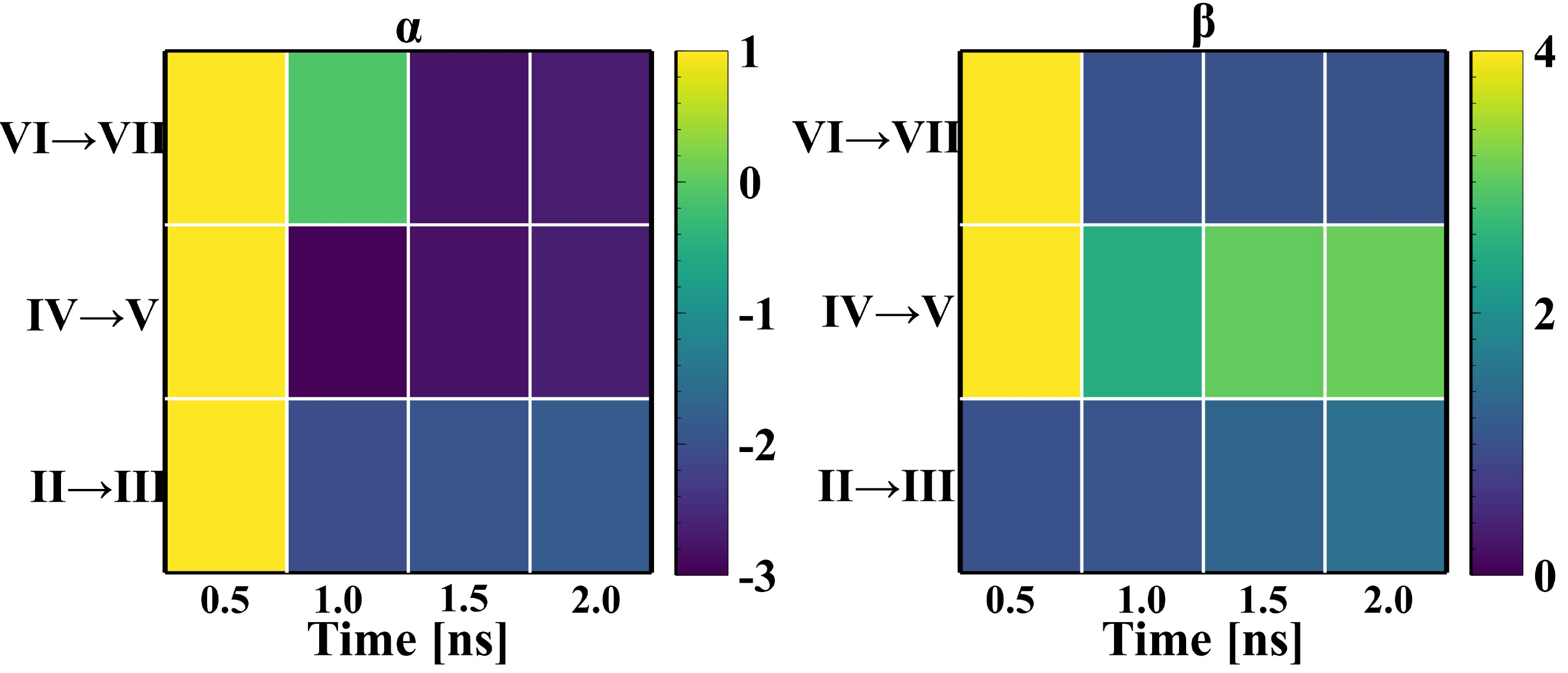}
\caption{ $\alpha$ (left panel) and $\beta$ (right panel) values for the nominal run. Time is given along the $x$ axis in units of ns. Transition is given along the $y$-axis. The value is for each ``pixel'' is given by the color bars. }
\label{fig:2DB100}
\end{center}
\end{figure*}

\begin{figure*}
\begin{center}
\includegraphics[trim=0.0mm 0.0mm 0.0mm 0.0mm, clip, scale=0.40]{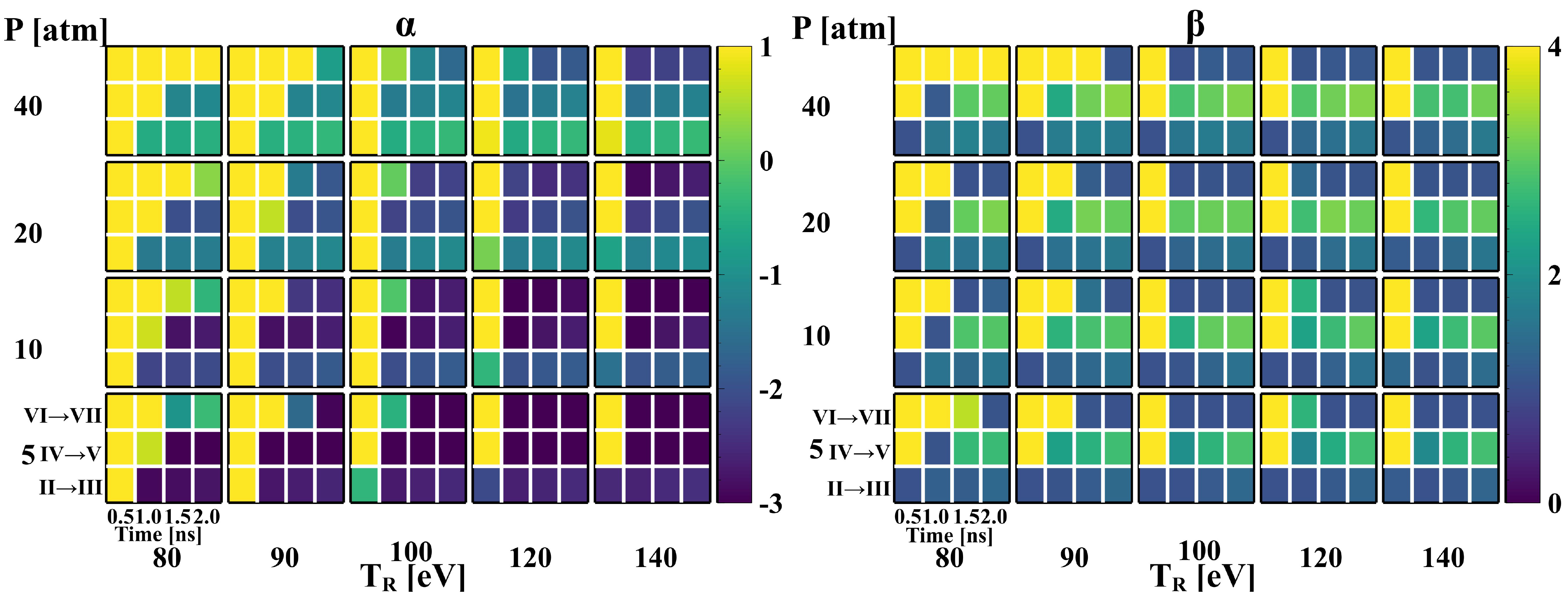}
\caption{ {\it Left Panel:} logarithmic $\alpha$ and {\it Right Panel:} $\beta$ values for the \omegafac\, models. A four by three panel represents each model. Pressure increases along the $y$-axis and the radiation source temperature increases along the $x$-axis. The time for each subplot is given along the $x$-axis in units of ns. Colorbars give the values for each cell. A photoionization front is described by an $\alpha\ll$1 and $\beta\sim$1. }
\label{fig:2DABOmega}
\end{center}
\end{figure*}

\begin{figure*}
\begin{center}
\includegraphics[trim=0.0mm 0.0mm 0.0mm 0.0mm,clip,scale=0.5]{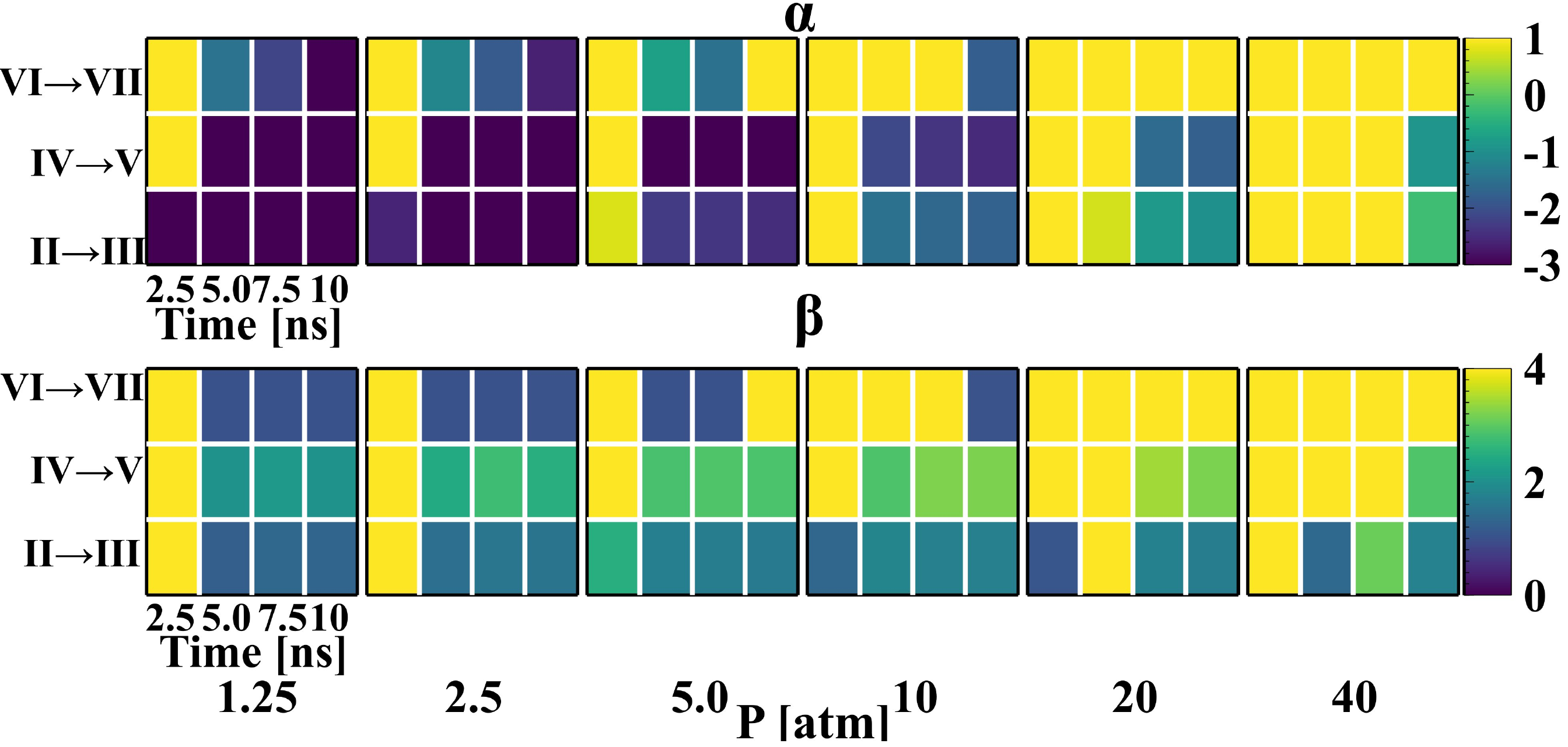}
\caption{{\it Top Panel: } logarithmic $\alpha$ and {\it Bottom Panel: } $\beta$ values for the ``z-pinch'' models. The four chosen times correspond to 2.5, 5.0, 7.5, and 10 ns. The initial nitrogen pressure increases along the $x$-axis. Colorbars give the values for each cell. }
\label{fig:2DABZ}
\end{center}
\end{figure*}

In this section, we turn our attention to determining whether a photoionization front is established in any of our simulations. There are two dimensionless quantities that are important in characterizing the photoionization front, the ratio of the recombination rate to the photoionization rate, denoted above as $\alpha_i$ (Eqn.~\ref{eqn:alpha1}), and the ratio of the coefficient of electron-impact (also known as collisional) ionization to that of recombination, denoted by $\beta_i$ (Eqn.~\ref{eqn:beta1}).

In order to estimate these ratios in our simulations we have taken the electron impact ionization rates from \cite{Voronov1997} and the recombination rate coefficients are computed using radiative recombination rate coefficients from \cite{Badnell2006} and the approximate three-body recombination rate coefficients from \cite{Lotz1967} and \cite{Drake2016}. \cite{Nikolic2013} showed that for high density plasmas, n$_{e}>$10$^{10}$ cm$^{-3}$, dielectronic recombination rate coefficients are highly suppressed. Since the densities studied here are much higher than this threshold, we assume that the contribution from dielectronic recombination is zero. We note here that these rate coefficients are not necessarily the same rates computed by {\sc Propaceos} and used in computing the opacity and equation of state tables used in the \crash\ models. However, these rates are sufficient in estimating $\alpha$ and $\beta$. The photoionization rate is approximated as, $\Gamma_i$ = $\bar{\sigma}_i \phi$, where $\bar{\sigma}_i$ is the average photoionization cross section for ion $i$ and $\phi$ is the radiation flux. The average photoionization cross section for ion $i$ is computed as
\begin{equation}
\label{eqn:sigmabar}
\bar{\sigma_i} =\frac{\int^{\infty}_{E_{th}} \sigma_i(\epsilon) F_{\epsilon}(T) d\epsilon}{\int^{\infty}_{0} F_{\epsilon}(T) d\epsilon} ,
\end{equation}
where $\sigma_i$ is the photoionization cross section for ion $i$, E$_{th}$ is the threshold energy for photoionization, and F$_{\epsilon}$ is the spectral photon flux. The photoionization cross sections are taken from \cite{Verner1995} and \cite{Verner1996}. The flux is computed as $\phi$=-${(c\lambda}/{\kappa})\nabla{E_R}$, where $c$ is the speed of light, E$_R$ is the radiation energy density, $\kappa$ is the Rosseland mean opacity, and the $\lambda$ is the flux limiter. Opacities are computed using a gray approximation for nitrogen, provided by {\sc Propaceos}.

The values of $\alpha$ and $\beta$ are computed for each ionization state of nitrogen along the line shown in Figure~\ref{fig:2DPlotsNorm}. For a given ionization state, we determine the location where X$_i$/X$_{i,max}$=0.5 which defines the photoionization front, using the (reasonably accurate) approximation that only two states are populated significantly at any one location. X$_i$(=$\rho_i/\rho$) is the mass fraction for state $i$ and is defined as the density of charge state $i$ divided by the total ion density and X$_{i,max}$ is the maximum value for ionization state $i$. X$_i$ is computed from the average ionization state as reported from the EOS tables which ranges between zero for completely neutral nitrogen and seven for fully ionized nitrogen. If the average ionization state is reported as Z=2.5, then the mass fraction for both NIII and NIV are set to 0.5. The mass fraction for each nitrogen ion is shown in Figure~\ref{fig:SpecTau} for the nominal model at t=1.5 ns. Note that the location where X$_i$/X$_{i,max}$=0.5 is multivalued and slightly complicates the determination of the photoionization front. Therefore, only the point closest to the radiation source is considered.

Figure~\ref{fig:FidTauPlot} shows the $\alpha$ and $\beta$ values for the nominal run at four times. At very early times (top left panel) only the first two ionization states of nitrogen are populated, but with very high $\alpha$ values, indicating a heat front and not a photoionization front. This is due to the relatively weak ionizing spectrum with a source temperature of T$_{\rm R}$=50 eV. However, after the source temperature has reached its peak value (t=1.0 ns) almost every ionization state is populated and, the parameters are suggestive of a photoionization front. For example, at t=1.5 ns only the VI$\rightarrow$V transition has a $\beta$ value appreciably above one. Error bars are also given in Figure~\ref{fig:FidTauPlot}. In most cases the $\alpha$ and $\beta$ values for the 25\% and 75\% ionization values do not differ from those at 50\%.

Another important feature of Figure~\ref{fig:FidTauPlot} is the location of each photoionization front as a function of time and optical depth. For example, at t=1.5 ns, there are effectively two or three groups of points. Each group roughly corresponds to an electron orbital subshell of nitrogen, with the 2p (outermost) subshell found at larger optical depths, then the 2s subshell, and the 1s (innermost) subshell appearing last. This confirms the results found in \S\ref{sec:threelevel} with multiple fronts being generated.

Three representative transitions are chosen to compare the results for the rest of the parameter study. II$\rightarrow$III, IV$\rightarrow$V, and VI$\rightarrow$VII are chosen as they represent transitions for each electron shell of nitrogen. As shown in Figure~\ref{fig:FidTauPlot}, IV$\rightarrow$V represents a worst-case in terms of $\beta$ for all the states considered, that is; the values for $\beta$ are consistently larger than other transitions especially at late times. Four representative times are chosen to compute the $\alpha$ and $\beta$ values for these transitions. The first time is chosen during the rise in source temperature (t=0.5 ns), another once the source temperature reaches its peak (t=1 ns), and two while the source temperature is held constant at the peak (t=1.5 and 2 ns).

Figure~\ref{fig:2DB100} shows the results of Figure~\ref{fig:FidTauPlot} in a slightly different manner, where each ``pixel'' gives the corresponding value for $\alpha$ and $\beta$. The interpretation is the same as in Figure~\ref{fig:FidTauPlot}. After the source has reached its peak value, the photoionization rate is higher than the recombination rate coefficients and gives values for $\alpha<<1$. Similarly, the electron collision ionization rate is smaller than the recombination rate and gives values for $\beta\sim$1.

Figure~\ref{fig:2DABOmega} shows the results for all the \omegafac-like runs where the nitrogen density varies from 2.5$\times$10$^{19}$ to 2$\times$10$^{21}$ cm$^{-3}$, corresponding to an initial pressure of 5 to 40 atmospheres respectively, and the peak radiation source temperature varies between 80 and 140 eV. A four by three panel similar to that of Figure~\ref{fig:2DB100}. represents each model. The source temperature T$_{\rm R}$ increases along the x-axis while the nitrogen pressure (alternatively density) increases along the y-axis.

As Figure~\ref{fig:2DABOmega} shows, photoionization fronts are generally expected for radiation source temperatures above 90 eV and nitrogen pressures between 5 and 10 atmospheres. Both of these constraints are readily achievable for current experimental setups. For laser driven experiments \cite{Davis2016} showed that radiation temperatures of 100 eV for 0.5 $\mu$m gold foils are possible, see Figure~\ref{fig:laserprofile}. Meanwhile, the ZAPP platform is well characterized in \cite{Rochau2014} and consistently produces 90 eV sources. In addition, the current experimental design makes use of a gas window that allows for gas pressures well within the ranges considered here. Therefore, it is likely that the photoionization fronts can be produced with the conditions considered here.

The right panel of Figure~\ref{fig:2DABOmega} suggests two important experimental considerations. First, at early times, $\beta$ is very large suggesting that these ionization states are populated through electron collisions instead of photoionizations. However, once the radiation source temperature has reached its peak, $\beta$ values are close to one over a wide range of source radiation temperatures and nitrogen pressures. One concludes that, data collected at early times, before 1 ns, is unlikely to see photoionization-dominated behavior. Secondly, as discussed above, the IV$\rightarrow$V transition is a worst-case scenario among all possible transitions but remains close to one, with peak values of $\beta\sim$3, suggesting that the photoionization front should be apparent for nearly every transition. Finally, this confirms the estimates made in \cite{Drake2016} and suggests that photoionization fronts should be produced with radiation temperatures of T$_{\rm R}>$90 eV and for nitrogen pressures between 5 and 20 atm.

Figure~\ref{fig:2DABZ} shows similar $\alpha$ (top panel) and $\beta$ (bottom panel) plots for the ``z-pinch'' models. The same atomic transitions are considered as in the \omegafac-like models. As shown in Figure~\ref{fig:laserprofile} the x-ray drive profile takes 10 ns to reach peak. Therefore, $\alpha$ and $\beta$ values are computed at 2.5, 5.0, 7.5, and 10 ns. Note due to the design of the ZAPP platform the radiation source temperature is fixed at T$_{\rm R}\sim$90 eV and only the initial nitrogen pressure is varied. The nitrogen pressure increases along the $x$ axis.

Lower nitrogen densities are favored in ``z-pinch'' experiments as compared to \omegafac-like experiments. At very high nitrogen pressure ( 5-10 atmospheres), both $\alpha$ and $\beta$ are very large which disfavors a photoionization front. In fact, a photoionization front is only favored at nitrogen pressures between 1.25 and 5 atmospheres. The understanding that optimum gas densities for the case of Z are roughly three times smaller than those for Omega suggests that an actual experiment for Z should be physically three times larger than the example shown in Figure~\ref{fig:setupZ}, if one desires for it to have the same scale in optical depths. Since the radiation drive takes time to develop, the photoionization front also takes some time to develop. In a real experiment, one would expect to observe photoionization fronts between 2.5 and 5.0 ns after the flux reaches the target. Similar to the \omegafac\, experiments, there is no preferred ionization state.

In this section, results from a parameter study looking for the formation of a photoionization front have been presented. Two dimensionless quantities are used to characterize the front $\alpha$, which relates electron recombinations to the photoionizing flux, and $\beta$, which relates the electron impact ionization to recombinations. A photoionization front is characterized by an $\alpha<<1$ and $\beta\sim1$. As shown in Figure~\ref{fig:2DABOmega} and Figure~\ref{fig:2DABZ} both the laser driven and ``z-pinch'' experiments are capable of generating true photoionization fronts. In fact, photoionization fronts are generated over a wide range of radiation source temperatures and nitrogen pressures.

\subsection{Comparison of Numerical and Theoretical Results}

We now compare the theoretical results and numerical results. With regard to the hydrodynamical model developed in \S\ref{sec:IonFronts}, we find very good agreement between the theory and simulations. For a weak "R" type front- the classic, simple photoionization front Figure~\ref{fig:mach}- shows the density is essentially unchanged. Figure~\ref{fig:lineout} shows that the density is nearly unchanged matching the theoretical predictions.

The three-level model of the atomic physics we developed in \S\ref{sec:threelevel} proved useful in interpreting the simulations. We introduced $\alpha$, the ratio of the electron recombination rate to the photoionization rate, and examined its relationship to the equilibrium ionization state fraction. As shown in Figure~\ref{fig:modelfrac} different values of $\alpha$ result in each ionization charge state dominating in distinct regions. This result is confirmed in Figure~\ref{fig:FidTauPlot}, which shows that each ionization state of nitrogen is found at distinct optical depths.

We note that although adding additional levels to the approximation presented in \S\ref{sec:threelevel} is straightforward, we find that three levels is sufficient our present purposes.

\section{Summary and Conclusion}

We have presented here both an improved theoretical model and a series of computational simulations, using a two-dimensional radiation hydrodynamics code, with the purpose of more accurately assessing whether feasible experiments can produce photoionization fronts. We have improved on the two-level model by formulating an approximate three-level model atom. Instead of a single value for $\alpha$ and $\beta$ an array of values are defined, one for each pair of adjacent charge states. The resulting ionization structure is strongly dependent on the values of $\alpha$, where similar values of $\alpha$ create very small regions where intermediate ions are dominant. If adjacent $\alpha$ values are vastly different, the intermediate ions exist over a range of optical depths. This suggests, therefore, that multiple ionization fronts can develop, one for each ionization state.

The simulations included the effects that could not be evaluated in the previous work using zero-dimensional models. These specifically include effects of electron heat conduction, of energy transport in two dimensions, and of the hydrodynamic response of the ionized gas. We conducted a parameter study in which we varied the temperature of the radiation source and pressure of the nitrogen gas. Two experimental platforms were studied, first a laser-driven platform in which a gold foil is irradiated by laser energy and creates the x-ray source, which is labeled as \omegafac\,-like; second a pulsed-power-driven platform in which the x-ray source is generated by the implosion of a current carrying wire, which is labeled as ``z-pinch'' and is based on the ZAPP platform \cite{Rochau2014}. Our nominal model is based on the ideal model in \citet{Drake2016} and uses the \omegafac-like platform.

We find that over a wide range of radiation temperatures and nitrogen pressures conditions consistent with the production of a photoionization front exist. In the \omegafac-like models, moderate nitrogen pressures and high radiation source temperatures are preferred. Specifically, initial nitrogen pressures between five and twenty atmospheres and radiation temperatures above 90 eV create ideal conditions. In the ``z-pinch'' models, nitrogen pressures between one and five atmospheres are ideal. We also find for our nominal model that electron heat conduction does not play a significant role in the forward propagation of the front. It is likely that heat conduction does play a role in the region where strong electron heating occurs, and in radial heat conduction behind the photoionization front.

The simulations presented here provide important insight into laboratory photoionization experiments. Importantly, two experimental platforms are studied and found to generate photoionization fronts. Within the limitations of our models, we conclude that the generation of photoionization fronts appears to be possible using present day experimental platforms using nitrogen gas. Within the context of the two-dimensional results shown here, one could better understand the implications of improved one-dimensional models of radiation transport and/or the atomic physics. This provides a direction for future research.

\acknowledgments
We would like to thank the anonymous referee for their comments, whose comments and suggestions greatly improved this paper. R.P.D and W.J.G were supported by the U.S. Department of Energy, through the NNSA-DS and SC-OFES Joint Program in High-Energy-Density Laboratory Plasmas, grant number DE-NA0002956, and by the Lawrence Livermore National Laboratory under subcontract B614207. The authors also acknowledge the Livermore Computing Center at Lawrence Livermore Nation Laboratory for providing HPC resources that contributed to the results reported within this paper.

\software{CRASH\citep{vanderHolst2011}, PROPACEOS\citep{MacFarlane2006}}

\bibliographystyle{apjsingle}
\bibliography{ms.bib}

\end{document}